\documentclass[final]{siamltex}

\usepackage{latexsym}
\usepackage{amssymb}
\usepackage{amsfonts}
\usepackage{graphicx}
\usepackage{subfigure}
\usepackage{lineno}
\usepackage{showkeys}

\newtheorem{remark}[theorem]{Remark}

\newcommand{\sech}{\rm{sech\,}}

\title{Breathing modes of long Josephson junctions with phase-shifts\thanks{Received by the editors \dots.}}

\author{
Amir Ali \thanks{School of Mathematical Sciences,
University of Nottingham, University Park, Nottingham, NG7 2RD, UK
({\tt amiralishahs@yahoo.com})} \and
Hadi Susanto \thanks{School of Mathematical Sciences,
University of Nottingham, University Park, Nottingham, NG7 2RD, UK
({\tt hadi.susanto@nottingham.ac.uk})} \and
Jonathan A.D.\ Wattis \thanks{School of Mathematical Sciences,
University of Nottingham, University Park, Nottingham, NG7 2RD, UK
({\tt jonathan.wattis@nottingham.ac.uk})}}

\begin{document}

\maketitle

\begin{abstract}
We consider a spatially inhomogeneous sine-Gordon equation with a time-periodic drive, modeling a microwave driven long Josephson junction with phase-shifts. Under appropriate conditions, Josephson junctions with phase-shifts can have a spatially nonuniform ground state. In recent reports [Phys. Rev. Lett. 98, 117006 (2007), arXiv:0903.1046], it is experimentally shown that a microwave drive can be used to measure the eigenfrequency of a junction's ground state. Such a microwave spectroscopy is based on the observation that when the frequency of the applied microwave is in the vicinity of the natural frequency of the ground state, the junction can switch to a resistive state, characterized by a non-zero junction voltage. It was conjectured that the process is analogous to the resonant phenomenon in a simple pendulum motion driven by a time periodic external force. In the case of long junctions with phase-shifts, it would be a resonance between the internal breathing mode of the ground state and the microwave field. Nonetheless, it was also reported that the microwave power needed to switch the junction into a resistive state depends on the magnitude of the eigenfrequency to be measured. 
Using multiple scale expansions, we show here that an infinitely long Josephson junction with phase-shifts cannot be switched to a resistive state by microwave field with frequency close to the system's eigenfrequency, provided that the applied microwave amplitude is small enough, which confirms the experimental observations. It is because higher harmonics with frequencies in the continuous spectrum are excited, in the form of continuous wave radiation. The breathing mode thus experiences  radiative damping. In the absence of driving, the breathing mode decays at rates of at most $\mathcal{O}(t^{-1/4})$ and $\mathcal{O}(t^{-1/2})$ for junctions with a uniform and nonuniform ground state, respectively. The presence of applied microwaves balances the nonlinear damping, creating a stable breather mode oscillation. As a particular example, we consider the so-called $0-\pi-0$ and $0-\kappa$ Josephson junctions, respectively representing the two cases. We confirm our analytical results numerically. Using our numerical computations, we also show that there is a critical microwave amplitude at which the junction switches to the resistive state. Yet, it appears that the switching process is not necessarily caused by the breathing mode. We show a case where a junction switches to a resistive state due to the continuous wave background becoming modulationally unstable.
\end{abstract}

\begin{keywords}
radiative damping, fractional fluxon, breathing modes, Josephson junction with phase-shifts, long Josephson junction.
\end{keywords}

\begin{AMS}
78A40, 34D05, 34D10, 34D20
\end{AMS}

\pagestyle{myheadings}
\thispagestyle{plain}
\markboth{ALI, SUSANTO, WATTIS}{Breathing modes of long Josephson junctions with phase-shifts}
\section{Introduction}
\label{intro}

A Josephson junction is an electronic circuit consisting of two superconductors connected by a thin nonsuperconducting layer, and is the basis of a large number of developments both in fundamental research and in application to electronic devices. Even though there is no applied voltage difference, a flow of electrons can tunnel from one superconductor to the other. This is due to the quantum mechanical waves in the two superconductors of the Josephson junction overlapping with each other. If we denote the difference in phases of the wave functions by $\phi$ and the spatial and temporal variable along the junction by $x$ and $t$, respectively, the electron flow tunneling across the barrier, i.e.\ the Josephson current, $I$   is proportional to the sine of $\phi(x,t)$, i.e.\ $I\sim\sin\phi(x,t)$. In a long Josephson junction, the phase difference $\phi$ satisfies a sine-Gordon equation.

In a standard long Josephson junction, the energetic ground state of the system is $\phi(x)$ constant satisfying $\sin\phi=\gamma$, where $\gamma$ is an applied constant (dc) bias current, which is taken to be zero here. A novel type of Josephson junction was proposed by Bulaevskii et al.\ \cite{bula77,bula78}, in which a non-trivial ground-state can be realized, characterized by the spontaneous generation of a fractional fluxon, i.e.\ a vortex carrying a fraction of magnetic flux quantum. This remarkable property can be invoked by intrinsically building  piecewise constant phase-shifts $\theta(x)$ into the junction. Due to the phase-shift, the supercurrent relation then becomes $I \sim\sin(\phi + \theta)$. Presently, one can impose a phase-shift in a long Josephson junction using several methods (see, e.g., \cite{hilg08,gold09} and references therein).

Due to this nontrivial properties, Josephson junctions with phase shifts may have promising applications in information storage and information processing \cite{pegr06,ortl06}. Because of their potential applications, the next natural question is: what is the eigenvalue of the ground state? It is important because Josephson junction based devices should not operate at frequencies close to the eigenfrequency of the system, as unwanted parasitic resonances can be induced. 

The eigenfrequency of the ground state in the simplest case of Josephson junctions with one and two phase-shifts has been theoretically calculated in \cite{kato97,gold05,ahma09,derk07,voge08,gold05_2}. More importantly, the eigenfrequency calculation in the former case has been confirmed experimentally recently in \cite{buck07,pfei09}. The experimental measurements were performed by applying  microwave radiation of fixed frequency and power to the Josephson junction. At some frequency, the junction interestingly switches to the resistive state, characterized by a non-zero junction voltage. In terms of the phase-difference $\phi$, the averaged Josephson voltage $<V>$ is proportional to 
\begin{equation}
<V>\sim\frac1T\int_{0}^T\int_{x\in\mathcal{D}}{\phi_t(x,t)\,dx\,dt},
\label{V}
\end{equation}
where $\mathcal{D}$ is the domain of the problem and $T\gg1$. It was conjectured that the  driving frequency at which switching occurs is the same as the eigenfrequency of the ground state \cite{buck07}. It is assumed that the jump to the resistive state is due to the resonant excitation of the breathing mode of the ground state and the applied microwaves, similar to the resonance phenomena observed in a periodically driven short (point-like) Josephson junction reported in \cite{gron04,gron04_2,guoz08}.

It was also noticed in \cite{buck07} that the accuracy of the microwave spectroscopy depends on the magnitude of the eigenfrequency. To measure a large natural frequency, the method requires an applied microwave with high power, which influences the measurement due to the nonlinearity of the system. Here, we consider an infinitely long Josephson junction with phase-shifts with no applied constant (dc) bias current. We show that in such a system, the breathing mode cannot be excited to switch the junction into a resistive state, provided that the microwave amplitude is small enough. It is the case even when the applied drive frequency is the same as the eigenfrequency, because of higher harmonic excitations in the form of continuous wave emission. In other words, the breathing mode experiences radiative damping. Such damping is not present in short junctions as the phase difference $\phi$ in that limit is effectively independent of $x$. This then confirms the observed experimental results.

The governing equation we consider herein is given by
\begin{equation}
\phi_{xx}(x,t)-\phi_{tt}(x,t)=\sin{\left(\phi+\theta(x)\right)}+h\cos(\Omega t),\quad x \in \mathbb{R},\,t>0,
\label{eq1}
\end{equation}
describing an infinitely long Josephson junction with phase-shifts driven by a microwave field. Equation (\ref{eq1}) is dimensionless, $x$ and $t$ are normalized by the Josephson penetration length $\lambda_J$ and the inverse plasma frequency $\omega_p^{-1}$ , respectively.
The applied time periodic (ac) drive in the governing equation above has amplitude $h$, which is proportional to the applied microwave power, and frequency $\Omega$. Here we study two cases of the internal phase-shift
\begin{equation}
\theta(x)=\left\{
\begin{array}{cc}
0,&|x|>a,\\
\pi, & |x|< a,
\end{array}
\right.
\label{th1}
\end{equation}
with $a<\pi/4$, and
\begin{equation}
\theta(x)=\left\{
\begin{array}{cc}
0,&x<0,\\
-\kappa, & x>0,
\end{array}
\right.
\label{th2}
\end{equation}
with $0< \kappa < 2\pi$, which are called $0-\pi-0$ and $0-\kappa$ Josephson junction, respectively. These are the simplest configurations admitting a uniform and a nonuniform ground state, respectively. The phase field $\phi$ is then naturally subject to the continuity conditions at the position of the jump in the Josephson phase (the discontinuity), i.e.\
\[
\phi(\pm a^-)=\phi(\pm a^+),\,\phi_x(\pm a^-)=\phi_x(\pm a^+),
\] for the $0-\pi-0$ junction and
\[
\phi(0^-)=\phi(0^+),\,\phi_x(0^-)=\phi_x(0^+),
\] for the $0-\kappa$ junction.

The unperturbed $0-\pi-0$ junction, i.e.\ (\ref{eq1}) and (\ref{th1}) with $h=0$, has
\[\Phi_0=0\] (mod 2$\pi$) as the ground state and by linearizing  around the uniform solution we find a breathing localized mode \cite{kato97}
\begin{equation}
\Phi_1(x,t)=e^{i\omega t}\left\{
\begin{array}{lll}
\cos(a\sqrt{1+\omega^2})e^{\sqrt{1-\omega^2}(a+x)},
& x< -a,\\
\cos(x\sqrt{1+\omega^2}),
 & \left|x\right| < a, \\
\cos(a\sqrt{1+\omega^2})e^{\sqrt{1-\omega^2}(a-x)},
 & x> a,
\end{array}\right.
\label{p11}
\end{equation}
with the oscillation frequency $\omega$ given by the implicit relation
\begin{eqnarray}
a=\frac {1}{\sqrt {1+{\omega}^{2}}} \tan^{-1} \left( \sqrt{\frac{1-{\omega}^{2}}{1 +{\omega}^{2}}} \right),\quad  \omega^{2}< 1.
\label{f1}
\end{eqnarray}

As for the unperturbed $0-\kappa$ junction, i.e.\ (\ref{eq1}) and (\ref{th2}) with $h=0$, the ground state of the system is  (mod $2\pi$)
\begin{eqnarray}
\Phi_0(x,t)=\left\{
\begin{array}{lll}
4\tan^{-1} e^{x_{0}+x},
& x<0, \\
\kappa-4\tan^{-1} e^{x_{0}-x},
 & x>0,
\end{array}
\right.
\label{p02}
\end{eqnarray}
where  $x_0=\ln\tan\left({\kappa}/{8}\right)$. Physically, $\Phi_0$ (\ref{p02}) represents a fractional fluxon that is spontaneously generated at the discontinuity. A scanning microscopy image of fractional fluxons can be seen in, e.g., \cite{hilg03,frol08}. Linearizing around the ground state $\Phi_0$ (\ref{p02}) we obtain the breathing mode  \cite{gold05,derk07}
\begin{eqnarray}
\Phi_{1}(x,t) = e^{i\omega t}\left\{
\begin{array}{lll}
e^{\Lambda(x_{0}+x)}\left[\tanh(x_{0}+x)-\Lambda \right],
& x<0, \\
e^{\Lambda(x_{0}-x)}\left[\tanh(x_{0}-x)-\Lambda \right],
 & x>0,
\end{array}
\right.
\label{p12}
\end{eqnarray}
with
\begin{eqnarray}
\Lambda=\sqrt{1-\omega^{2}},
\end{eqnarray}
and the oscillation frequency
\begin{eqnarray}
\omega(\kappa)=\pm\sqrt{\frac{1}{2}\cos\frac{\kappa}{4}\left(\cos\frac{\kappa}{4}+\sqrt{4-3\cos^2\frac{\kappa}{4}}\,\right)}.
\label{f2}
\end{eqnarray}
In addition to the eigenfrequency (\ref{f1}) or (\ref{f2}), a ground state in a Josephson junction also has a continuous spectrum in the range $\omega^2>1.$

\begin{figure}[h]
  \begin{center}
    \includegraphics[width=0.65\textwidth,angle=0]{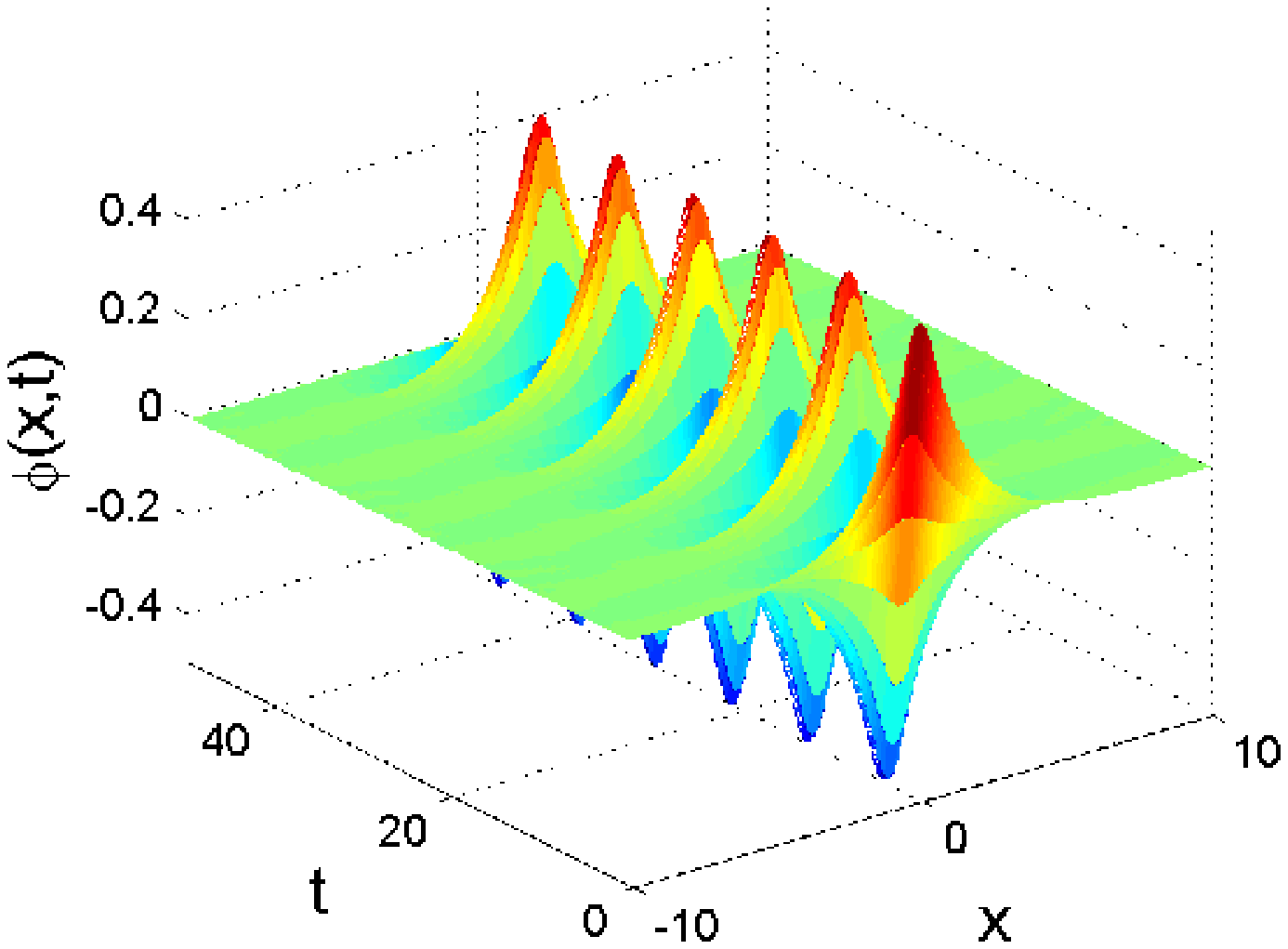}
    \includegraphics[width=0.65\textwidth,angle=0]{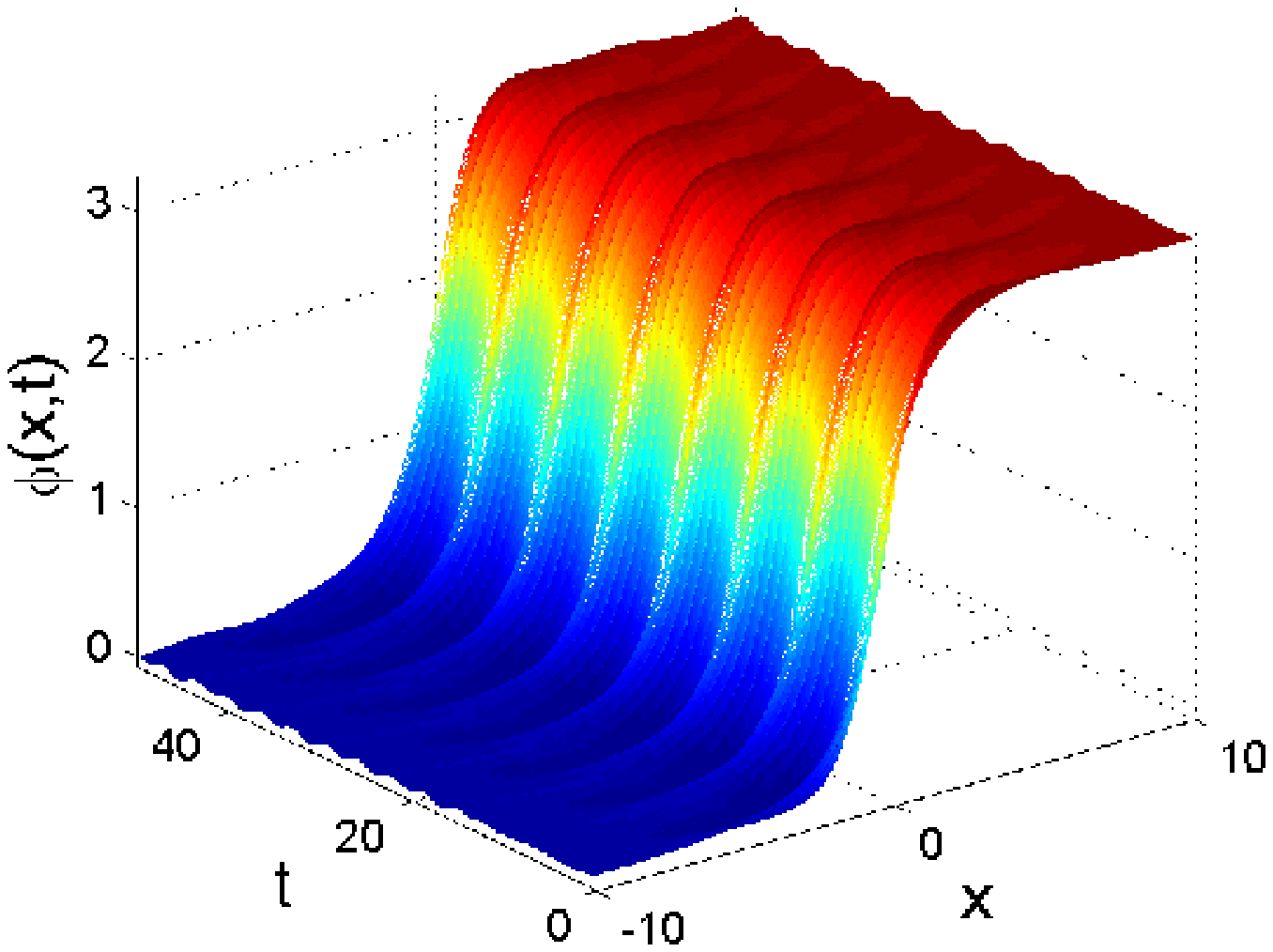}
  \end{center}
  \caption{The typical  dynamics of a breathing mode (top) and a wobbling kink (bottom) in an undriven $0-\pi-0$ and $0-\kappa$ junction, respectively.}
  \label{fig1}
\end{figure}

If a ground state is perturbed by its corresponding localized mode, then the perturbation will oscillate periodically . The typical  evolution of the initial condition
\begin{equation}
\phi=\Phi_0(x)+B_0\Phi_1(x,0),
\label{init}
\end{equation}
for some small initial amplitude $B_0$ and $h=0$ is shown in the top and bottom panel of Fig.\ \ref{fig1} for the two cases above. For a $0-\pi-0$ junction, one can see a clear mode oscillation on top of the uniform background state $\phi=0$. In the case of a $0-\kappa$ junction, the periodic oscillation of the localized mode makes the fractional kink oscillate about the point of discontinuity,  $x=0$.

Using a multiple scale expansion, we show that in the absence of an ac drive, such a breathing mode oscillation decays with a rate of at most $\mathcal{O}(t^{-1/4})$ and $\mathcal{O}(t^{-1/2})$ for a junction with a uniform and nonuniform ground state, respectively.
The coupling of a spatially localized breathing mode to radiation modes via a nonlinearity and the same decay rates have been discussed and obtained by others in several contexts (see \cite{oxto09} and references therein). Interactions of a breathing mode and a topological kink, creating the so-called 'wobbling kink', or simply 'wobbler', have also been considered  before, see \cite{oxto09,oxto09_2} (see also references therein) for $\phi^4$ wobblers and \cite{segu83,kalb04,fere08} for sine-Gordon wobblers. Nonetheless, the problem and results presented herein are novel and important from several points of view, which include the fact that our fractional wobbling kink is in principal different from the `normal' wobbler. Usually, a wobbler is a periodically expanding and contracting kink, due to the interaction of the kink and its \emph{odd} eigenmode. Because our system is not translationally invariant, our wobbler will be composed of a fractional kink and an \emph{even} eigenmode, representing a topological excitation oscillating about the discontinuity point (see also \cite{kevr00} for a similar situation in discrete systems, where a lattice kink interacts with its even mode). Such an oscillation can certainly be induced by a time periodic direct driving, as considered herein. More importantly, our problem is relevant and can be readily confirmed experimentally (see also, e.g., \cite{dewe08,{hans06},gold04} for experimental fabrications of $0-\pi-0$ Josephson junctions).

The presentation of the paper is outlined as follows. In Section 2, we construct a perturbation expansion for the breathing mode to obtain equations for the slow-time evolution of the oscillation amplitude in a $0-\pi-0$ junction by eliminating secular terms from our expansion. In Section 3, the method of multiple scales is applied to obtain the amplitude oscillation in the presence of driving, extending the preceding section. In Sections 4 and 5, we apply the perturbation method to the wobbling kink in a $0-\kappa$ junction. We confirm our analytical results numerically in Section 6. We also show in the same section that there is a threshold drive amplitude above which the junction switches to the resistive state. Yet, we observe that the switching to the resistive state is due to the modulational instability of the background. We conclude the present work in Section 7. 

\section{ Freely oscillating breathing mode in a $0-\pi-0$ junction}

In this section we construct a breathing mode of the sine-Gordon equation (\ref{eq1}) with $h=0$ and $\theta$ given by (\ref{th1}).

We apply a perturbation method to the equation (\ref{eq1}) by writing
\begin{eqnarray}
\phi=\phi_{0}+\epsilon\,\phi_{1}+\epsilon^2\phi_{2}+\epsilon^3\phi_{3}+\dots,
\end{eqnarray}
where \ $\epsilon  $ is a small parameter. We further use multiple scale expansions by introducing the slow time and space variables
\begin{eqnarray}
 X_{n}=\epsilon^{n}x, \quad  T_{n}=\epsilon^{n}t, \quad n=0,1,2,\dots,
\label{mt}
 \end{eqnarray}
which describes long times and distances. In the small limit of $\epsilon$, the different scales above become uncoupled and may be considered as independent variables. 

In the following, we  use the notation
\begin{eqnarray}
\partial_{n}=\frac{\partial}{\partial X_n}, \quad D_n=\frac{\partial}{\partial T_n},
\label{a5}
\end{eqnarray}
such that the derivatives with respect to the original variables in terms of the scaled variables using the chain rule are given by
 \begin{eqnarray}
&&\frac{\partial}{\partial{x}} = \partial_0+\epsilon\,\partial_1+\epsilon^2\partial_2+\epsilon^3\partial_3+\dots,\\
&&\frac{\partial}{\partial{t}}= D_0+\epsilon\, D_1+\epsilon^2D_2+\epsilon^3D_3+\dots.
\end{eqnarray}

Substituting these expansions into the perturbed sine-Gordon equation (\ref{eq1}) along with the expansion of $\phi$ and equating like powers of $\epsilon $, we obtain a hierarchy of linear partial differential equations:
\begin{eqnarray}
&&\mathcal{O}(1):\,\partial_0^2\phi_0-D_0^2\phi_0=\sin(\theta+\phi_0),\\
&&\mathcal{O}(\epsilon):\,\partial_0^2\phi_1-D_0^2\phi_1-\cos(\theta+\phi_0)\phi_1=2D_0D_1\phi_0-2\partial_0\partial_1\phi_0.
\end{eqnarray}
Solutions to the equations above for the $0-\pi-0$ junction are given by
\begin{eqnarray}
\phi_0(X_0,T_0)=0,
\end{eqnarray}
and
\begin{eqnarray}
\phi_1(X_0,T_0)=B(X_1,..,T_1,..)\Phi_1(X_0,T_0)+c.c.,
\end{eqnarray}
where $\Phi_{1}$ is given by (\ref{p11}). $B(X_{1},\dots,T_{1},\dots)$ is the amplitude of the breathing mode, which is a function of the slow time and space variables only. Throughout the paper, $c.c.$ stands for the complex conjugate of the immediately preceding term.

Next, we consider the $\mathcal{O}(\epsilon^2)$ equation
\begin{eqnarray}
\partial_0^2\phi_2- D_0^2 \phi_2 - \cos(\theta\ + \phi_{0})\phi_2 =2D_0D_1\phi_1 - 2\partial_0\partial_1\phi_1.
\label{p21}
\end{eqnarray}
Evaluating the right hand side for the different regions, we obtain
\begin{eqnarray*}
\partial_0^2\phi_2-D_0^2\phi_2-\phi_2&=&2\cos(a\sqrt{1+\omega^2})\left[i\omega D_{1} B-\sqrt{1-\omega^2}\partial_1 B\right]e^{\sqrt{1-\omega^2}(a+X_0)+i\omega T_0},\\
\partial_0^2\phi_2-D_0^2\phi_2+\phi_2&=&2\left[i\omega D_1B \cos(X_{0}\sqrt{1+\omega^2}) +\sqrt{1+\omega^2}\partial_1B \sin(X_{0}\sqrt{1+\omega^2})\right]e^{i\omega T_0},\\
\partial_0^2\phi_2-D_0^2\phi_2-\phi_2&=&2\cos(a\sqrt{1+\omega^2})\left[i\omega D_1B+\sqrt{1-\omega^2} \partial_1B\right] e^{\sqrt{1-\omega^2}(a-X_0)+i\omega T_0},
\end{eqnarray*}
for $X_0<-a$, $|X_0|<a$, and $X_0>a$, respectively. These are linear wave equations with forcing having frequency $ \omega $.\\

Substituting the spectral ansatz $\phi_2(X_0,T_0)=\tilde{\phi_2}(X_0)e^{i\omega T_0}$, we obtain the corresponding set of ordinary differential equations with forcing term which has the frequency $\omega$
\begin{eqnarray*}
\partial_0^2\tilde{\phi_2}-(1-\omega^2)\tilde{\phi_2}&=&2\cos(a\sqrt{1+\omega^2})\left[i\omega D_{1}B-\sqrt{1-\omega^2}\partial_{1}B\right]e^{\sqrt{1-\omega^2}(a +X_{0})},\\
\partial_0^2\tilde{\phi_2}+(1+\omega^2)\tilde{\phi_2}&=&2\left[i\omega D_{1}B\cos(X_{0}\sqrt{1+\omega^2}) +\sqrt{1+\omega^2}\partial_{1} B\sin(X_{0}\sqrt{1 +\omega^2})\right],\\
\partial_0^2\tilde{\phi_2}-(1-\omega^2)\tilde{\phi_2}&=&2\cos(a\sqrt{1+\omega^2})\left[i\omega D_{1}B+\sqrt{1-\omega^2}\partial_{1}B\right] e^{\sqrt{1-\omega^2} (a-X_0)}.
\end{eqnarray*}
We need to find the bounded solution of the above equations, which are of the form $T\psi\left(X_0\right)=f\left(X_0\right)$ where $T$ is a self-adjoint operator given by the left hand side of the above system.

The Fredholm theorem states that the necessary and sufficient condition for the above nonhomogeneous equation to have a bounded solution is that its right-hand side  $f(X_0)$  be orthogonal to the complete system of linearly independent solutions of the corresponding homogeneous equation, $T\psi(X_0)=0$. By applying the theorem, we find the solvability condition
\begin{equation}
D_1B=0.
\label{t1}
\end{equation}

The bounded solution of (\ref{p21}) is then given by
\begin{eqnarray*}
\phi_2=\partial_{1}B e^{i\omega T_{0}}\left\{\begin{array}{lll}
{{e}^{\sqrt {1-{\omega}^{2}}X_{0}}}{\it C_{21}}-X_{0}\cos \left( a\sqrt {1+{\omega}^{2}} \right){{e}^{\sqrt {1-{\omega}^{2}}\left( a+X_{0}\right) }}+c.c.,
&  X_0< -a,\\
\cos \left(X_{0}\sqrt {1+{\omega}^{2}}\right) {\it C_{22}}-X_{0}\cos \left( X_{0}\sqrt {1+{\omega}^{2}} \right)+c.c.,
 & \left|X_0\right| < a, \\
{{e}^{-\sqrt {1-{\omega}^{2}}X_{0}}}{\it C_{23}}-X_{0}\cos \left( a\sqrt {1+{\omega}^{2}} \right){{e}^{\sqrt {1-{\omega}^{2}}\left( a-X_{0} \right) }}+c.c.,
 & X_0> a,
\end{array}\right.
\end{eqnarray*}
where $C_{21}=C_{23}$ and $C_{22}$ are constants of integration that have to be found by applying the continuity conditions at the discontinuity points $X_0=\pm a$.

It should be noted that $ \partial_{1}B $ as well as $ \partial_{n}B $ in later calculations do not appear in the solvability condition. Therefore, we take the simplest choice by setting \[ \partial_{1}B =0.\]
This choice is also in accordance with the fact that if $\partial_{1}B$ was non zero, then $(\epsilon^{2}\phi_{2})$ would become greater than $(\epsilon \phi_{1})$ as $X_0\to\pm\infty$ due to the term $(X_{0}e^{\sqrt{1-{\omega}^{2}}\left( a\mp X_{0}\right)})$ in the expression of $\phi_2$ above, leading to a nonuniformity in the perturbation expansion of $\phi$.

The equation at the third order in the perturbation expansion is
\begin{eqnarray}
\partial_{0}^{2}\phi_{3}-D_{0}^{2}\phi_{3}-\cos(\theta)\phi_{3}=2(D_{0}D_{2}
-\partial_{0}\partial_{2}) \phi_{1}+(D_{1}^{2}-\partial_{1}^{2})\phi_{1}-\frac{1}{6}\phi_{1}^{3}\cos(\theta).
\label{a11}
 \end{eqnarray}
Having evaluated the right hand side using the functions $\phi_{0}$ and $\phi_{1}$, and splitting the solution into components proportional to simple harmonics we get
\begin{eqnarray*}
\partial_{0}^{2}\phi_{3}-D_{0}^{2}\phi_{3}-\cos(\theta)\phi_{3}=\left\{\begin{array}{lll}
F_{1},
&  X_0< -a,\\
F_{2},
 & \left|X_0\right| < a, \\
F_{3},
 &X_0> a,
\end{array}\right.\nonumber
\end{eqnarray*}
where $F_{1}$, $F_{2}$, $F_{3}$ are given by
\begin{eqnarray*}
F_{1}&=&2\left(i\omega D_{2}B-\sqrt {1-\omega^{2}}\partial_{2}B\right)\cos \left(a\sqrt {1+\omega^{2}}\right)e^{\sqrt {1-{\omega}^{2}}(a+X_{0})+i\omega T_{0}}\\
&&-\frac{1}{2}B|B|^{2}\cos^{3}\left(a\sqrt {1+{\omega}^{2}}\right)e^{3\sqrt {1-{\omega}^{2}}\left(a+X_{0}\right)+i\omega T_{0}}\\
&&-\frac{1}{6} B^{3}\cos^{3} (a\sqrt {1+{\omega}^{2}})e^{3\sqrt {1-{\omega}^{2}}( a+X_{0})+{3i\omega\,T_{0}}},\\
F_{2}&=&\left[2\,i\omega D_{2}B\cos\left( \sqrt {1+{ \omega}^{2}}X_{0}\right)+2\partial_{2}B\,\sqrt {1+{\omega}^{2}}\sin\left(\sqrt{1+{\omega}^{2}} X_{0}\right)\right.\\
&&+ \left.\frac{1}{2}B|B|^{2} \cos^{3} \left(\sqrt{1+{\omega}^{2}}X_{0}\right)\right] e^{i\omega T_{0}}+\frac{1}{6}B^{3}\cos^{3}\left(\sqrt {1+{\omega}^{2}}X_{0}\right)e^{3i\omega T_{0}},\\
F_{3}&=&2\left(i\omega D_{2}B+\sqrt {1-\omega^{2}}\partial_{2}B\right)\cos \left(a\sqrt {1+\omega^{2}}\right)e^{\sqrt {1-{\omega}^{2}}(a-X_{0})+i\omega T_{0}}\\
&&-\frac{1}{2}B|B|^{2}\cos^{3}\left(a\sqrt {1+{\omega}^{2}}\right)e^{3\sqrt {1-{\omega}^{2}}\left(a-X_{0}\right)+i\omega T_{0}}\\
&&-\frac{1}{6} B^{3}\cos^{3} (a\sqrt {1+{\omega}^{2}})e^{3\sqrt {1-{\omega}^{2}}( a-X_{0})+{3i\omega\,T_{0}}}.
\end{eqnarray*}
These are linear wave equations with forcing at frequencies $\omega$ and $3\omega $. The former frequency is resonant with the discrete eigenmode and the latter is assumed to lie in the continuous spectrum (phonon band), i.e.\
\begin{equation}
9\omega^2>1.
\label{3w1}
\end{equation}

As (\ref{a11}) is linear, the solution can be written as a combination of solutions with frequencies present in the forcing, i.e.
\begin{eqnarray}
\phi_3=\phi_3^{(0)}+\phi_3^{(1)}e^{i\omega T_0}+c.c.+\phi_3^{(2)}e^{2i\omega T_0}+c.c.+\phi_3^{(3)}e^{3i\omega T_0}+c.c.
\label{har1}
\end{eqnarray}
This implies that $\phi_3^{(1)}$ satisfies the following nonhomogeneous equations in the three regions below
\begin{eqnarray}
\partial_{0}^{2}\phi_{3}^{(1)}-\left(\cos(\theta)-\omega^{2}\right)\phi_{3}^{(1)}=\left\{\begin{array}{lll}
2\,i\omega\,{\it D_{2}B}\,\cos \left( a\sqrt {1+{\omega}^{2}} \right) {e^{\sqrt {1-{\omega}^{2}} \left( a+X_{{0}} \right) }}\\-\frac{1}{2}\,B|B|^{2}  \cos^{3} \left(a \sqrt {1+{\omega}^{2}}\right){e^{3\,\sqrt {1-{\omega}^{2}} \left( a+X_{{0}} \right) }},
&   X_0< -a,\\
2\,i\omega\,{\it D_{2}B}\cos \left(\sqrt {1+{\omega}^{2}}X_{{0}}\right)\\
+\frac{1}{2}\,B|B|^{2} \cos^{3}\left(\sqrt {1+{\omega}^{2}}X_{{0}}\right),
&  \left|X_0\right| < a, \\
2\,i\omega\,{\it D_{2}B}\,\cos \left(a \sqrt {1+{\omega}^{2}}\right) {e^{\sqrt {1-{\omega}^{2}} \left( a-X_{{0}} \right) }}\\
-\frac{1}{2}\,B|B|^{2} \cos^{3} \left( a\sqrt {1+{\omega}^{2}}\right){e^{3\,\sqrt {1-{\omega}^{2}} \left( a-X_{{0}} \right)}},
&X_0> a.
\end{array}\right.\nonumber
\end{eqnarray}
In the above equation, it should be noted that we have imposed the fact that \[\partial_{2}B =0,\] as we previously discussed.

The solvability condition for the first harmonic gives
\begin{eqnarray}
D_{2}B=i k_{1}B|B|^{2},
\label{eq2}
\end{eqnarray}
where
\begin{eqnarray}
k_{1}=\frac{\left( 3-7\,{\omega}^{4}-2\,{\omega}^{6}-2\,{\omega}^{2}+6\,\sqrt {1-{\omega}^{4}}\tan^{-1} \left( \sqrt{\frac{1-{\omega}^{2}}{1 +{\omega}^{2}}} \right)  \right)}{32\,\omega\, \left( 1+{\omega}^{2}+\sqrt {1-{\omega}^{4}} \tan^{-1} \left( \sqrt{\frac{1-{\omega}^{2}}{1 +{\omega}^{2}}} \right) \right)}.
\label{a6}
\end{eqnarray}
The solution for the first harmonic is then given by
\begin{eqnarray}
\phi_{3}^{(1)}(X_{0},T_{0})=B|B|^{2}\left\{\begin{array}{lll}
\upsilon_{1}(X_{0}),
&  X_0< -a,\\
\upsilon_{2}(X_{0}),
 &  \left|X_0\right| < a, \\
\upsilon_{3}(X_{0}),
 & X_0> a,
\end{array}\right.
\label{phi31_0p0}
\end{eqnarray}
where
\begin{eqnarray*}
&&\upsilon_{1}(X_{0})={{e}^{\sqrt {1-{\omega}^{2}}X_{{0}}}}{\it C_{31}}-\frac{\sqrt {2\,(1+{\omega}^{2})} \left( k_{{31}}{{e}^{2\,\sqrt {1-{\omega}^{2}}X_{{0}}}}-\tilde{k_{{31}}} \right){{e}^{-\sqrt {1-{\omega}^{2}}X_{{0}}}}}{64\,\omega\, \sqrt {1-{\omega}^{2}}\,u_{1} }\\
&&\upsilon_{2}(X_{0})={\rm Re} ({{e}^{i\,\sqrt {1+{\omega}^{2}}X_{{0}}}}) {C_{32}}-\frac{\tilde{k_{{32}}}{\rm Re}({{e}^{i\,\sqrt {1+{\omega}^{2}}X_{{0}}}})-k_{{32}}{\rm Im} ({{e}^{i\,\sqrt {1+{\omega}^{2}}X_{{0}}}})}{16\,\sqrt {1+{\omega}^{2}} u_{2}},\\
&&\upsilon_{3}(X_{0})={{e}^{-\sqrt {1-{\omega}^{2}}X_{{0}}}}{\it C_{33}}-\frac{ \left( k_{{33}}{{e}^{2\,\sqrt {1-{\omega}^{2}}X_{{0}}}}-\tilde{k_{{33}}} \right){{e}^{-\sqrt {1-{\omega}^{2}}X_{{0}}}}}{32\,\sqrt {(2-{\omega}^{2})({1-{\omega}^{2})}}\,u_{1} },
\end{eqnarray*}
with
\begin{eqnarray*}
u_{1}=\left( 1+{\omega}^{2}+\sqrt {1-{\omega}^{4}}\tan^{-1} \sqrt{\frac{1-{\omega}^{2}}{1+{\omega}^{2}}}\right), \quad u_{2}=\left( {\omega}^{2}+\sqrt {1-{\omega}^{4}}\tan^{-1}  \sqrt {1-{\omega}^{4}}\right).
\end{eqnarray*}
The expression of the functions ${k_{{3j}}}$ and $\tilde{k_{{3j}}}$, $j=1,2,3,$ is given in (\ref{k31})--(\ref{k33t}). The coefficients $ C_{31}=C_{32} $ and $ C_{33} $ are constants of integration, that should be determined from the continuity conditions at $ X_{0}=\pm a $.

We do not consider the equation for the second harmonic $\phi_3^{(2)}$ as it does not appear in the leading order of the sought asymptotic equation describing the behavior of breathing mode amplitude.

The equation for the third harmonic $\phi_3^{(3)}$ is
\begin{eqnarray*}
\partial_{0}^{2}\phi_{3}^{(3)}- \left(\cos(\theta)-9\omega^{2}\right)\phi_{3}^{(3)}=\left\{\begin{array}{lll}
-\frac{1}{6} \cos^{3} ( \sqrt {1+{\omega}^{2}}a)e^{3\sqrt {1-{\omega}^{2}}( a+X_{0})} ,
&  X_0< -a,\\
\frac{1}{6}\cos^{3}( \sqrt {1+{\omega}^{2}}X_{0}),
&\left|X_0\right| < a, \\
-\frac{1}{6}\cos^{3} ( \sqrt {1+{\omega}^{2}}a) e^{3\sqrt {1-{\omega}^{2}}( a-X_{0})},
& X_0> a,
\end{array}\right.\nonumber
\end{eqnarray*}
whose solution,  using the same procedure as above, is given by
\begin{eqnarray*}
\phi_{3}^{(3)}\left(X_{0},T_{0}\right)=B^{3} \left\{\begin{array}{lll}
P_{31}(X_{0}),
& X_0< -a,\\
P_{32}(X_{0}),
& \left|X_0\right| < a, \\
P_{33}(X_{0}),
 &  X_0> a,
\end{array}\right.
\end{eqnarray*}
where
\begin{eqnarray*}
P_{31}(X_{0})&=&{{e}^{\sqrt {1-9\,{\omega}^{2}}X_{{0}}}}\tilde{C_{{31}}}-\frac{1}{48}\,  \cos^{3}\left( a\sqrt {1+{\omega}^{2}}\right) {{e}^{3\,\sqrt {1-{\omega}^{2}} \left( a+X_{{0}} \right) }}, \\
P_{32}(X_{0})&=&\cos(\sqrt {1+9\,{\omega}^{2}}X_{{0}}) {\it \tilde{C_{32}}}  -\frac{1}{192{\,\omega}^{2}}\left({\omega}^{2} -3\right)\,\cos \left( X_{{0}}\sqrt {1+{\omega}^{2}} \right), \\
P_{33}(X_{0})&=&{{e}^{-\sqrt {1-9\,{\omega}^{2}}X_{{0}}}}\tilde{C_{{33}}}-\frac{1}{48}\, \cos^{3}\left( a\sqrt {1+{\omega}^{2}}\right ) {{e}^{3\,\sqrt {1-{\omega}^{2}} \left( a-X_{{0}} \right) }},
\end{eqnarray*}
and
$\tilde{C_{{31}}}$, $\tilde{C_{{32}}}$, and $\tilde{C_{{33}}}$ are non-zero constants of integration that also are determined from the continuity conditions at the discontinuity points.

Note that due to the assumption (\ref{3w1}), the second term in $P_{31}(X_{0})$ and $P_{33}(X_{0})$ will decay to zero while the first term tends to $\tilde{C_{{31}}}$ and $\tilde{C_{{33}}}$ as $X_0\to\mp\infty$, respectively. This implies that $e^{3i\omega T_0}\phi_{3}^{(3)}+c.c.$ represents a continuous wave radiation traveling to the left and right.

Solving the $\mathcal{O}\left(\epsilon^{4}\right)$  equation we obtain
\[\phi_{4}=0,\]
from the solvability condition
\begin{equation}
D_3B=0,
\label{t2}
\end{equation}
which is similar to the case of $\phi_{2}$.

Equating terms at $\mathcal{O}(\epsilon^{5})$ gives the equation
\begin{eqnarray*}
{\partial_{{0}}}^{2}\phi_{{5}}-{D_{{0}}}^{2}\phi_{{5}}-\cos ( \theta )\phi_{{5}}=2(D_{0}D_{4}-\partial_{0}\partial_{4})\phi_{1}+2(D_{3}D_{1} -\partial_{3}\partial_{1})\phi_{1} +(D^{2}_{2}-\partial^{2}_{2})\phi_{1}\\+(D_{{1}}^{2}-\partial_{{1}}^{2})\phi_{3}  +2(D_{{2}}D_{{0}} -\partial_{{2}}\partial_{{0}})\phi_{3} +\left(-\frac{1}{2}\,{\phi_{{1}}}^{2}\phi_{{3}}+{\frac {1}{120}}{\phi_{{1}}}^{5} \right)\cos (\theta ).
\end{eqnarray*}
Having calculated the right hand side using the known functions, we again split the solution into components proportional to simple harmonics as we did before. The equation for the first harmonic is given by
\begin{eqnarray}
\partial_{0}^{2}\phi_{5}^{(1)}- \left(\cos(\theta)-\omega^{2}\right)\phi_{5}^{(1)}=\left\{\begin{array}{lll}
G_{1} ,
&  X_0< -a,\\
G_{2},
 & \left|X_0\right| < a, \\
G_{3},
 & X_0> a,
\end{array}\right.
\label{a1}
\end{eqnarray}
where
\begin{eqnarray*}
G_{1}&=&2\,i\omega\,D_{{4}}B\cos\left( a\sqrt {1+{\omega}^{2}}\right) {{e}^{\sqrt {1-{\omega}^{2}} ( a+X_{{0}}) }}\\&&-B|B|^4 \left[ {k_{{1}}}^{2} \cos\left( a\sqrt {1+{\omega}^{2}}\right ) {{e}^{\sqrt {1-{\omega}^{2}}( a+X_{{0}}) }}+2\omega k_{{1}} \upsilon_{1}( X_{{0}})\right.\\&&\left. +\frac{1}{2}\,\cos  ^{2}\left( a\sqrt {1+{\omega}^{2}}\right){{e}^{2\,\sqrt {1-{\omega}^{2}}( a+X_{{0}}) }}( 3\upsilon_{1}( X_{{0}}) +P_{{31}}( X_{{0}}) )\right.\\ &&
\left.-\frac{1}{12}\, \cos^{5}\left( a\sqrt {1+{\omega}^{2}}\right) {{e}^{5\,\sqrt {1-{ \omega}^{2}}( a+X_{{0}}) }}\right],\\
G_{2}&=&2\,i\omega\,D_{{4}}B\cos\left( X_{{0}}\sqrt {1+{\omega}^{2}}\right )- B|B|^4\left[ {k_{{1}}}^{2}\cos \left( X_{{0}}\sqrt {1+{\omega}^{2}}\right ) +2\omega k_{{1}}\upsilon_{2}( X_{{0}})\hspace*{4 cm} \right.\\&&\left.-\frac{1}{2}\, \cos^{2}\left( X_{{0}}\sqrt {1+{\omega}^{2}}\right)  ( 3\,\upsilon_{2}( X_{{0}}) +P_{{32}} ( X_{{0}} )) +\frac{1}{12}\, \cos^{5}\left ( X_{{0}}\sqrt {1+{\omega}^{2}}\right)\right ], \\
G_{3}&=&2\,i\omega\,D_{{4}}B\cos \left( a\sqrt {1+{\omega}^{2}}\right) {{e}^{\sqrt {1-{\omega}^{2}} ( a-X_{{0}}) }}\\&&-B|B|^4\left[{k_{{1}}}^{2}\cos \left( a\sqrt {1+{\omega}^{2}} \right) {{e}^{\sqrt {1-{\omega}^{2}}( a-X_{{0}}) }}+2\omega k_{{1}}\upsilon_{3}( X_{{0}})\right.\\&& \left.+\frac{1}{2}\, \cos^{2} \left( a\sqrt {1+{\omega}^{2}}\right) {{e}^{2\,\sqrt {1-{\omega}^{2}}( a-X_{{0}}) }}( 3\,\upsilon_{3}( X_{{0}}) +P_{{33}} ( X_{{0}}) )\right.\\&& \left. -\frac{1}{12}\,\cos^{5} \left( a\sqrt {1+{\omega}^{2}}\right) {{e}^{5\,\sqrt {1-{\omega}^{2}} ( a-X_{{0}}) }}\right].
\end{eqnarray*}
Here, $\upsilon_{{1}}( X_{{0}})$, $\upsilon_{{2}}( X_{{0}})$, $\upsilon_{{3}} \left( X_{{0}} \right)$ are the bounded solutions of $\phi^{(1)}_{3}(X_{0},T_{0})$ and $P_{{31}}( X_{{0}})$, $P_{{32}} ( X_{{0}})$, $P_{{33}}( X_{{0}})$ are the bounded solutions of $\phi^{(3)}_{3}(X_{0},T_{0})$ as solved above.

The solvability condition of (\ref{a1}) is
\begin{eqnarray}
D_{4}B=k_{2}B|B|^{4},
\label{eq3}
\end{eqnarray}
where
\begin{eqnarray*}
k_{2}&=&-\frac{\Upsilon_{2}\,i}{2\omega \Psi(\omega)}\in \mathbb{C},\quad \Upsilon_{2}=k^{2}_{1}\Psi(\omega)+2 \omega k_{1}\zeta(X_{0})+\alpha(X_{0}) +\beta(X_{0})+\gamma(X_{0}),\\
\Psi(\omega)&=&\frac{ \left( \sqrt {1+{\omega}^{2}}+\sqrt {1-{\omega}^{2}}\tan^{-1} \left( \sqrt{\frac{1-{\omega}^{2}}{1+{\omega}^{2}}} \right)\right)}{\sqrt{1-\omega^{4}}},\\
\zeta(X_{0})&=&\int _{-\infty }^{-a}\!\upsilon_{1}(X_{{0}}) \cos \left( a\sqrt {1+{\omega}^{2}}\right) {{e}^{\sqrt {1-{\omega}^{2}}( a+X_{{0}} ) }}{dX_{{0}}}+\int _{-a}^{a}\!\upsilon_{2}(X_{{0}})\cos \left( X_{{0}}\sqrt {1+{\omega}^{2}}\right){dX_{{0}}}\\
&&+\int _{a}^{\infty }\!\upsilon_{3}(X_{{0}})\cos \left( a\sqrt {1+{\omega}^{2}}\right) {{e}^{\sqrt {1-{\omega}^{2}} ( a-X_{{0}} )}} {dX_{{0}}},\\
\alpha(X_{0})&=&\frac{1}{2}\int _{-\infty }^{-a}\!\left( 3\,\upsilon_{1}( X_{{0}}) +P_{{31}}( X_{{0}})\right)\cos ^{3}\left( a\sqrt {1+{\omega}^{2}}\right) {{e}^{3\,\sqrt {1-{\omega}^{2}}( a+X_{{0}}) }} {dX_{{0}}}-\,{\frac { \cos^{6}( a\sqrt {1+{\omega}^{2}})}{72\sqrt {1-{\omega}^{2}}}},\\
\beta(X_{0})&=&-\frac{1}{2}\int _{-a}^{a}\!\left( 3\,\upsilon_{2}( X_{{0}}) +P_{{32}} ( X_{{0}})\right) \cos^{3}\left( X_{{0}}\sqrt {1+{\omega}^{2}}\right) {dX_{{0}}}+\frac{1}{12}\int _{-a}^{a}\!\cos^{6} ( X_{{0}}\sqrt {1+{\omega}^{2}}) {dX_{{0}}},\\
\gamma({X_{0}})&=&\frac{1}{2}\int _{a}^{\infty }\!\left( 3\,\upsilon_{3} \left( X_{{0}} \right) +P_{{33}}( X_{{0}})\right )\cos^{3}\left( a\sqrt {1+{\omega}^{2}}\right) {{e}^{3\,\sqrt {1-{\omega}^{2}}( a-X_{{0}} ) }} {dX_{{0}}}-{\frac {\cos^{6} ( a\sqrt {1+{\omega}^{2}} )}{72\sqrt {1-{\omega}^{2}}}}.
\end{eqnarray*}

We shall postpone the continuation of the perturbation expansion to higher orders as we have obtained the oscillating and decaying behavior of the breathing amplitude (\ref{eq2}) and (\ref{eq3}), which is our main objective.

By noting that \begin{eqnarray*}
\frac{d B}{d{t}}=\epsilon D_1B+\epsilon^2 D_2B+\epsilon^3 D_3B+\epsilon^4 D_4B,
\end{eqnarray*}
and defining $b=\epsilon\, B$, (\ref{t1}), (\ref{eq2}), (\ref{t2}), and (\ref{eq3}) can be combined to obtain
\begin{eqnarray}
\frac{db}{dt}=k_{1}b|b|^{2}\,i+k_{2}b|b|^{4}.
\label{a2}
\end{eqnarray}
One can calculate that the solution of (\ref{a2}) satisfies the relation
\begin{eqnarray}
|b(t)|=\left(\frac{|b(0)|^{4}}{1-4\,{\rm Re}\left( k_{2}\right)|b(0)|^{4}t}\right)^{\frac{1}{4}},
\label{a3}
\end{eqnarray}
where $b(0)$  is the initial amplitude of oscillation. By assuming that ${\rm Re}( k_{2})<0$, as will be shown later, this equation describes the gradual decrease in the amplitude of the breathing mode with order $\mathcal{O}({t^{-1/4}})$ as it emits energy in the form of radiation. 

\begin{remark}
The $\mathcal{O}\left({t^{-1/4}}\right)$ decay of the oscillation amplitude is because of our assumption (\ref{3w1}). If one has $\left(3\omega\right)^2<1$ instead, then the decay rate will be smaller than $\mathcal{O}\left({t^{-1/4}}\right)$ as the coefficient $k_{2}$ in (\ref{eq3}) will be purely imaginary. This then leads us to the following conjecture.
\begin{proposition}
If $n\geq3$ is an odd integer such that
\[
1/(n-2)^2>\omega^2\geq1/n^2,
\]
then the decay rate of the breathing mode oscillation in $0-\pi-0$ Josephson junctions with $a<\pi/4$ is of order $\mathcal{O}(t^{-1/(n+1)})$.
\end{proposition}

This proposition implies that the closer the eigenfrequency $\omega$ to zero, i.e.\ $a\to\pi/4$, the longer the life time of the breathing mode oscillation.
\label{r1}
\end{remark}



\section{Driven breathing mode in $0-\pi-0$ junction}

We now consider breathing mode oscillations in a $0-\pi-0$ junction in the presence of external driving with frequency near the natural breathing frequency of the mode, i.e.\ (\ref{eq1}) and (\ref{th1}) with $h\neq0$ and $\Omega=\omega(1+\rho)$.

By rescaling the time $\Omega t=\omega\tau$, (\ref{eq1}) becomes
\begin{eqnarray}
\phi_{xx}(x,\tau)-(1+\rho)^{2}\phi_{\tau\tau}(x,\tau)=\sin{(\phi+\theta)}+\frac12h\left(e^{i\omega \tau}+c.c.\right).
\label{a4}
\end{eqnarray}
Here, we assume that the driving amplitude and frequency are small, namely
\begin{equation}
h=\epsilon^{3} H, \,\rho=\epsilon^{3} R,
\end{equation}
with $H, R \sim \mathcal{O}(1)$.

Due to the time rescaling above, our slow temporal variables are now defined as
\begin{eqnarray}
 X_{n}=\epsilon^{n}x, \quad  T_{n}=\epsilon^{n}\tau, \quad n=0,1,2,\dots.
 \end{eqnarray}
In the following, we still use the short hand notations (\ref{a5}).

Performing a perturbation expansion order by order as before, one  obtains  the same perturbation expansion up to and including order $\epsilon^{2}$  as in the undriven case above. At  third order, we obtain
\begin{eqnarray}
\hspace*{1 cm}\partial_{0}^{2}\phi_{3}-D_{0}^{2}\phi_{3}-\cos(\theta)\phi_{3}&=&(D_{1}^{2}-\partial_{1}^{2})\phi_{1} +2(D_{0}D_{2} -\partial_{0}\partial_{2})\phi_{1}-\frac{1}{6}\phi_{1}^{3}\cos(\theta)\\&&+\frac12H\left(e^{i\omega \tau}+c.c.\right)\nonumber.
\end{eqnarray}
The only difference from the undriven case is the presence of a harmonic drive in the last term.

The first harmonic component of the above equation  gives us the solvability condition
\begin{eqnarray}
D_{2}B=k_{1} B|B|^{2}\,i+l_{1} H\,i,
\label{eq5}
\end{eqnarray}
where
\begin{eqnarray}
l_{1}=\frac{ \sqrt {1+{\omega}^{2}}}{\sqrt {2}\,\omega \left( 1+{\omega}^{2}+\sqrt {1-{\omega}^{4}}\tan^{-1} \left( \sqrt{\frac{1-{\omega}^{2}}{1 +{\omega}^{2}}} \right)\right)},
\label{a13}
\end{eqnarray}
and $k_1$ is given in (\ref{a6}). The solution for the first harmonic can then be readily obtained as
\begin{eqnarray*}
\phi_{3}^{(1)}=\left\{\begin{array}{lll}
B|B|^2\upsilon_{1}(X_{0})+H\tilde{\upsilon_{1}}(X_{0}),
&  X_0< -a,\\
B|B|^2\upsilon_{2}(X_{0})+H\tilde{\upsilon_{2}}(X_{0}),
 & \left|X_0\right| < a, \\
B|B|^2\upsilon_{3}(X_{0})+H\tilde{\upsilon_{3}}(X_{0}),
 & X_0> a,
\end{array}\right.\nonumber
\end{eqnarray*}
where
\begin{eqnarray*}
\tilde{\upsilon_{1}}(X_{0})&=&{{e}^{\sqrt {1-{\omega}^{2}}X_{{0}}}}{\tilde{\aleph}_{{31}}}-\frac{\sqrt {1+{\omega}^{2}} \left( 1-{\omega}^{2} \right)\tan^{-1} \left( \sqrt{\frac{1-{\omega}^{2}}{1 +{\omega}^{2}}} \right) }{2\,\left( 1-{\omega}^{2} \right) ^{3/2}u_{1}}\label{v4}\\
&&\frac{\,{{e}^{u+X_{{0}}}}\left((\omega^{4}-1)X_{{0}}+\sqrt {1-{\omega}^{2}}(1+\omega^{2})\right)-2\,\sqrt {1-{\omega}^{2}}(1+{\omega}^{2})}{4\,\left( 1-{\omega}^{2} \right) ^{3/2}u_{1}} ,\nonumber\\
\tilde{\upsilon_{2}}(X_{0})&=&{\rm Re}({e}^{i\,\sqrt {1+{\omega}^{2}}X_{{0}}}){\tilde{\aleph}_{{32}}}+\frac{1}{2\,( 1+{\omega}^{2})}\label{v5}\\&&
-\frac{\sqrt{1+\omega^{2}}X_{{0}}{\rm Im}\,({e}^{i\,\sqrt {1+{\omega}^{2}}X_{{0}}}) +{\rm Re}({e}^{i\,\sqrt {1+{\omega}^{2}}X_{{0}}})}{\sqrt {2}\, \sqrt{1+{\omega}^{2}}\,u_{2}},\nonumber\\
\tilde{\upsilon_{3}}(X_{0})&=&{{e}^{-\sqrt {1-{\omega}^{2}}X_{{0}}}}{\tilde{\aleph}_{{33}}}-\frac{\tan^{-1} \left( \sqrt{\frac{1-{\omega}^{2}}{1 +{\omega}^{2}}} \right)}{2\,u_{1}  }\label{v6}\\&&+\frac{{{e}^{-\sqrt {1-{\omega}^{2}}X_{{0}}}}(({{e}^{u}}- 2{{e}^{\sqrt {1-{\omega}^{2}}X_{{0}}}})\sqrt{1-{\omega}^{2}} ( 1+{\omega}^{2})+2\,{{e}^{u}}X_{{0}}( 1-{\omega}^{4})) }{4\,(1-{\omega}^{2}) ^{3/2}\,u_{1}}\nonumber ,
\end{eqnarray*}
and $u=\sqrt {1-{\omega}^{2}}\tan^{-1} \left( {\frac {\sqrt {1-{\omega}^{4}}}{1+{\omega}^{2}}} \right)$. The constants of integration $\tilde{\aleph}_{31}=\tilde{\aleph}_{33}$ and $\tilde{\aleph}_{{32}}$ are determined by applying the continuity conditions at the discontinuity points.

One can check that the solution for the third harmonic $\phi_{3}^{(3)}(X_{0},T_{0})$ is the same as in the undriven case as well as $\phi_{3}^{(2)}$, which has no role in the leading order asymptotic expansion. 

The equation at $\mathcal{O}({\epsilon^4})$ is
\begin{eqnarray*}
{\partial_{{0}}}^{2}\phi_{{4}}-{D_{{0}}}^{2}\phi_{{4}}-\cos \left(\theta+\phi_{0} \right) \phi_{{4}}=2\, \left( D_{{0}}D_{{1}}-\partial_{{0}}\partial
_{{1}} \right) \phi_{{3}}+2\, \left( D_{{1}}D_{{2}}+D_{{0}}D_{{3}} \right) \phi_{{1}}\\-2\, \left( \partial_{{1}}\partial_{{2}}+\partial_{{0}}
\partial_{{3}} \right) \phi_{{1}}+4\,R{D_{{0}}}^{2}\phi_{{1}},
\end{eqnarray*}
with the solvability condition for the above equation
\begin{eqnarray*}
D_{3}B=-2i\omega BR.
\end{eqnarray*}
This implies that $\phi_{4}=0$ as for the case of $\phi_{2}$.

At $\mathcal{O}(\epsilon^{5})$, we obtain
\begin{eqnarray*}
{\partial_{{0}}}^{2}\phi_{{5}}-{D_{{0}}}^{2}\phi_{{5}}-\phi_{{5}}\cos ( \theta )=2(D_{0}D_{4}-\partial_{0}\partial_{4})\phi_{1}+2(D_{3}D_{1} -\partial_{3}\partial_{1})\phi_{1} +(D^{2}_{2}-\partial^{2}_{2})\phi_{1}\\+(D_{{1}}^{2}-\partial_{{1}}^{2})\phi_{3}+2(D_{{2}}D_{{0}} -\partial_{{2}}\partial_{{0}})\phi_{3} -\left(\frac{1}{2}\,{\phi_{{1}}}^{2}\phi_{{3}}-{\frac {1}{120}}{\phi_{{1}}}^{5}\right )\cos (\theta )+4\,R\,D_{0}D_{1}\phi_{1}.
\end{eqnarray*}
Evaluating the right hand side, we again split the solution into components proportional to simple harmonics as we did before. For the first harmonic, we obtain that
\begin{eqnarray}
\partial_{0}^{2}\phi_{5}^{(1)}- \left(\cos(\theta)-\omega^{2}\right)\phi_{5}^{(1)}=\left\{\begin{array}{lll}
M_{1} ,
&  X_0< -a,\\
M_{2},
 & \left|X_0\right| < a, \\
M_{3},
 & X_0> a,
\end{array}\right.
\label{a7}
\end{eqnarray}
where
\begin{eqnarray*}
M_{1}&=&\left(2\,i\omega\,{\it D_{4}B}-{k_{1}}^{2}B|B|^{4}-k_{1}l_{1}{|B|}^{2}H\right)\cos\left( a\sqrt {1+{\omega}^2}\right) {{e}^{\sqrt{1-{\omega}^{2}}\left(a+X_{0}\right)}}\\
&&- 2\,\omega\,\left(k_{{1}}B{|B|}^{4}+l_{{1}}{|B|}^{2}H\right)\upsilon_{{1}}( X_{{0}})+\frac{1}{12}\,B{|B|}^{4} \cos^{5}\left( a\sqrt {1+{\omega}^{2}}\right){{e}^{5\,\sqrt {1-{\omega}^{2}}(a+X_{{0}})}}\\&& -\frac{1}{2}\,B{|B|}^{4} \left( 3\,\upsilon_{{1}}\left( X_{{0}}\right)+P_{{31}} \left( X_{{0}}\right)\right )\cos^{2}\left( a\sqrt {1+{\omega}^{2}}\right){{e}^{2\,\sqrt {1-{\omega}^{2}}\left(a+X_{{0}}\right)}}\\
&&-\frac{1}{2}\,H\left( 2\,{|B|}^{2} +{B}^{2}\right) \tilde{\upsilon}_{{1}}\left( X_{{0}}\right)\cos ^{2}\left(a\sqrt {1+{\omega}^{2}} \right){{e}^{2\,\sqrt {1-{\omega}^{2}}\left( a+X_{{0}}\right) }}, \\
M_{2}&=& \left( 2\,i\omega\,{\it D_{4}B}-{k_{{1}}}^{2}B|B|^{4}-k_{{1}}l_{{1}}|B|^{2}H \right) \cos \left( X_{{0}}\sqrt {1+{\omega}^{2}}\right)\\&& -2\,\omega\, \left( k_{{1}}B|B|^{4}+l_{{1}}|B|^{2}H \right) \upsilon_{{2}}\left( X_{{0}}\right)-\frac{1}{12}\,B|B|^{4} \cos^{5} \left( X_{{0}}\sqrt {1+{\omega}^{2}}\right)\\&& +\frac{1}{2}\,B|B|^{4}\left( 3\,\upsilon_{{2}} \left( X_{{0}}\right ) +P_{{32}}\left ( X_{{0}}\right )\right ) \cos ^{2}\left( X_{{0}}\sqrt {1+{\omega}^{2}}\right) \\&& +\frac{1}{2}\,H \left( 2\,|B|^{2} +{B}^{2}\right)\tilde{\upsilon_{2}}( X_{{0}} )\cos^{2} \left( X_{{0}}\sqrt {1+{\omega}^{2}}\right ) ,\\
M_{3}&=&\left( 2\,i\omega\,{\it D_{{4}}B}-{k_{1}}^{2}B{|B|}^{4}-k_{1}l_{1}{|B|}^{2}H \right)\cos\left( a\sqrt {1+{\omega}^2}\right) {{e}^{\sqrt{1-{\omega}^{2}}(a-X_{0})}}\\
&&-2\,\omega\,\left(k_{{1}}B{|B|}^{4}+l_{{1}}{|B|}^{2}H\right) \upsilon_{{3}}( X_{{0}})+\frac{1}{12}\,B{|B|}^{4} \cos^{5}\left ( a\sqrt {1+{\omega}^{2}}\right) {{e}^{5\,\sqrt {1-{\omega}^{2}}\left( a-X_{{0}}\right) }}\\&& -\frac{1}{2}\,B{|B|}^{4}\left( 3\,\upsilon_{{3}} ( X_{{0}}) +P_{{33}} ( X_{{0}}) \right)  \cos^{2} \left( a\sqrt {1+{\omega}^{2}}\right){{e}^{2\,\sqrt {1-{\omega}^{2}}( a-X_{{0}}) }}\\
&&-\frac{1}{2}\,H\left( 2\,{|B|}^{2} +{B}^{2}\right) \tilde{\upsilon}_{{3}}( X_{{0}})\cos ^{2}\left ( a\sqrt {1+{\omega}^{2}} \right){{e}^{2\,\sqrt {1-{\omega}^{2}}\left( a-X_{{0}}\right) }}.
\end{eqnarray*}
The solvability condition for (\ref{a7}) is
\begin{eqnarray}
D_{4}B=k_{2}\, B|B|^{4}+{H i}\left(l_2|B|^{2}+l_3B^{2}\right),
\label{eq6}
\end{eqnarray}
where $l_{1}$ is given by (\ref{a13}) and
\begin{eqnarray*}
l_2&=&\frac{{\Upsilon}_{3,1}}{{2\omega \Psi(\omega)}},\,l_3=\frac{{\Upsilon}_{3,2}}{{2\omega \Psi(\omega)}},\\
{\Upsilon}_{3,1}&=&k_{1}l_{1}\Psi(\omega)+2\omega l_{1}\zeta(X_{0}) +2\left(\alpha_{2}(X_{0}) +\beta_{2}(X_{0}) +\gamma_{2}(X_{0})\right),\\
{\Upsilon}_{3,2}&=&\alpha_{2}(X_{0})+\beta_{2}(X_{0})+\gamma_{2}(X_{0}),\\
\alpha_{2}(X_{0})&=&\frac{1}{2}\int _{-\infty }^{-a}\tilde{\upsilon_{1}}( X_{{0}} ) \cos^{3}\left ( a\sqrt {1+{\omega}^{2}}\right) {{e}^{3\,\sqrt {1-{\omega}^{2}}\left( a+X_{{0}} \right) }}{dX_{{0}}},\\
\beta_{2}(X_{0})&=&-\frac{1}{2}\int _{-a}^{a}\tilde{\upsilon_{2}}( X_{{0}} )  \cos ^{3}\left( X_{{0}}\sqrt{1+{\omega}^{2}}\right){dX_{{0}}},\\
\gamma_{2}(X_{0})&=&\frac{1}{2}\int _{a}^{\infty }\tilde{\upsilon_{3}}( X_{{0}} ) \cos^{3} \left( a\sqrt {1+{\omega}^{2}}\right) {{e}^{3\,\sqrt {1-{\omega}^{2}}( a-X_{{0}} ) }}{dX_{{0}}}.\hspace*{4.5 cm}
\end{eqnarray*}
So far we have obtained the sought leading order behavior of the breathing amplitude. Performing the same calculation as in (\ref{a2}), we obtain the governing dynamics of the oscillation amplitude in the presence of an external drive
\begin{eqnarray}
\frac\omega\Omega\frac{db}{dt}=k_{1} b|b|^{2}\,i+l_{1}h\,i+k_{2}\,b|b|^{4}+{h}\left(l_2|b|^{2}+l_3b^{2}\right)\,i-2\,i\,\omega\,b\,\rho.
\label{a8}
\end{eqnarray}
One can deduce that a nonzero external driving can induce a breathing dynamic. It is expected that for large $t$, there will be a balance between the external drive and the radiation damping.

\section{ Freely oscillating breathing mode in $0-\kappa $ junction}
In this section, we will consider (\ref{eq1}) with $\theta$ given by (\ref{th2}), describing the dynamics of the Josephson-phase in the $0-\kappa$ long Josephson junction.

By applying the method of multiple scales and the perturbation expansion as before to the governing equation, we obtain from the leading order $\mathcal{O}(1)$ and $\mathcal{O}(\epsilon)$ that
\begin{eqnarray}
\phi_{0}=\Phi_0(X_0), \quad \phi_{1}= B(X_1..,T_1..)\Phi_1(X_0,T_0)+c.c.,
\end{eqnarray}
where $\Phi_0$  and $\Phi_1$ are given by (\ref{p02}) and (\ref{p12}), respectively.

Using the fact that $\phi_0$ is a function of $X_0$ only, the equation at $\mathcal{O}(\epsilon^2)$ is
\begin{eqnarray}
\partial_0^2\phi_2-D_0^2\phi_2-\cos(\theta+\phi_0)\phi_2=2D_0D_1\phi_1-2\partial_0\partial_1\phi_1-\frac{\phi_1^2}{2} \sin(\theta+\phi_0).
\end{eqnarray}

After a simple algebraic calculation, one can recognize that the right hand side of the above equation consists of functions having frequencies $0 $,  $ \omega $ and $ 2 \omega $. Therefore, solutions to the above equation can be written as
\begin{eqnarray}
\phi_2=\phi_2^{(0)}+\phi_2^{(1)}e^{i\omega T_0}+c.c.+\phi_2^{(2)}e^{2i\omega T_0}+c.c.,
\end{eqnarray}
which implies that $\phi_2^{(0)}$, $\phi_2^{(1)}$ and $\phi_2^{(2)}$ satisfy the following nonhomogeneous equations in the two  regions $X_0<0$ and $X_0>0$
\begin{eqnarray*}
&&\Big(\partial_0^2-\cos(\theta+\phi_0)\Big)\phi^{(0)}_2=-{\left|B\right|}^{2}\left\{
\begin{array}{lll}
e^{2\sqrt{1-\omega^{2}}(x_0+X_{0})}\Big[\tanh(x_0+X_{0})-\sqrt{1-\omega^{2}}\Big]^2\sin(\phi_0),
& X_{0}<0, \\
e^{2\sqrt{1-\omega^{2}}(x_0-X_{0})}\Big[\tanh(x_0-X_{0})-\sqrt{1-\omega^{2}}\Big]^2\sin(\phi_0-\kappa),
 & X_{0}>0,
\end{array}
\right.\\
&&\Big(\partial_0^2+\omega^{2}-\cos(\theta+\phi_0)\Big)\phi^{(1)}_2=\left\{
\begin{array}{lll}
2\Big[\left(i\omega D_1B -\partial_{1}B\sqrt{1-\omega^{2}}\right)\left(\tanh(x_0+X_{0})-\sqrt{1-\omega^{2}}\right)\\
-\partial_{1}B\,{\rm sech}^{2}( x_{0}+X_{0})\Big]e^{\sqrt{1-\omega^{2}}(x_0+X_{0})},
&X_{0}<0, \\
2\Big[\left(i\omega D_1B +\partial_{1}B\sqrt{1-\omega^{2}}\right)\left(\tanh(x_0-X_{0})-\sqrt{1-\omega^{2}}\right)\\
+\partial_{1}B\,{\rm sech}^{2}( x_{0}-X_{0})\Big]e^{\sqrt{1-\omega^{2}}(x_0-X_{0})},
 & X_{0}>0,
\end{array}
\right.\\
&&\Big(\partial_0^2+4\omega^{2}-\cos(\theta+\phi_0)\Big)\phi^{(2)}_{2}=-\frac{B^2}{2}\left\{
\begin{array}{lll}
e^{2\sqrt{1-\omega^{2}}(x_{0}+X_{0})}\Big[\tanh(x_0+X_{0})-\sqrt{1-\omega^{2}}\Big]^{2}\sin(\phi_0),
& X_{0}<0, \\
e^{2\sqrt{1-\omega^{2}}(x_{0}-X_{0})}\Big[\tanh(x_0-X_{0})-\sqrt{1-\omega^{2}}\Big]^{2}\sin(\phi_0-\kappa),
 & X_{0}>0.
\end{array}
\right.
\end{eqnarray*}
By using arguments as in the preceding sections, we set
\[D_1B=0,\quad \partial_{1}B =0.\]
Hence, we find that
\[\phi^{(1)}_{2}(X_0,T_0)=0.\]
The solution for the other harmonics are
\begin{eqnarray}
\phi^{(0)}_{2} &=&|B|^{2}\left\{
\begin{array}{cc}
E_{0}(X_{0}),
& X_{0}<0, \\
\tilde{E_{0}}(X_{0}),
 & X_{0}>0,
\end{array}
\right.
\label{phi20_0k}
\\
\phi^{(2)}_{2} &=& B^{2}\left\{
\begin{array}{cc}
E_{2}(X_{0}),
& X_{0}<0, \\
\tilde{E_{2}}(X_{0}),
 & X_{0}>0,
\end{array}
\right.\label{phi22_0k}
\end{eqnarray}
where $E_{0}(X_{0})$, $\tilde{E_{0}}(X_{0})$,$E_{2}(X_{0})$ and $\tilde{E_{2}}(X_{0})$ are given in (\ref{v7})--(\ref{v10}). $C_{01}$, $C_{02}$, $C_{{21}}$ and $C_{{22}}$ are constants of integration that should be found by applying continuity condition at the point of discontinuity $X_0=0$.

In the following, we will assume that the harmonic $2\omega$ is in the continuous spectrum, i.e.\
\begin{equation}
4\omega^2>1.
\label{2w1}
\end{equation}
With this assumption, $ \phi^{(2)}_{2}(X_{0},T_{0})$ will not decay in space and $e^{2i\omega T_0} \phi^{(2)}_{2}(X_{0},T_{0})$ describes right moving radiation for positive $X_{0}$ and left moving radiation for negative $X_{0}$.

 Equating the terms at $\mathcal{O}(\epsilon^3)$ gives the equation
\begin{eqnarray*}
\partial_0^2\phi_3-D_0^2\phi_3-\cos(\theta+\phi_0)\phi_3&=&(2D_0D_2 -2\partial_0\partial_2)\phi_1+(D_1^2-\partial_1^2)\phi_1+\nonumber\\
&&(2D_0D_1 -2\partial_0\partial_1)\phi_2 -\phi_1\phi_2\sin(\theta+\phi_0)\\&&-\frac{\phi_1^3}{6}\cos(\theta+\phi_0),
\end{eqnarray*}
where we have used the fact that $\phi_0$ depends only on $X_0$. Having calculated the right hand side using the known functions $\phi_{0}$, $\phi_{1}$ and $\phi_{2}$, we again split the solution into components proportional to the harmonics of the right hand side. Specifically for the first harmonic, we have
\begin{eqnarray*}
\Big(\partial_{0}^{2}+\omega^{2}-\cos\left(\theta+\phi_{0}\right)\Big)\phi^{(1)}_{3} =
\left\{\begin{array}{cc}
L_{1},
& X_{0}<0, \\
L_{2},
& X_{0}>0,
\end{array}
\right.
\end{eqnarray*}
where
\begin{eqnarray*}
L_{1}&=&2i\omega\,{ D_{2}B}\left[\tanh ( { x_{0}}+X_{{0}}) -\sqrt {1-{\omega}^{2}}\right]{{e}^{\sqrt {1-{\omega}^{2}}({ x_{0}}+X_{{0}}) }} \\&&-B|B|^2\left( {E}_{{0}} ( X_{{0}}) +{E}_{{2}} ( X_{{0}}) \right)\sin(\phi_0)\left[\tanh ( { x_{0}}+X_{{0}}) -\sqrt {1-{\omega}^{2}}\right]{{e}^{\sqrt {1-{\omega}^{2}}({ x_{0}}+X_{{0}}) }} \\&& -\frac{1}{2}{B|B|^2}\cos(\phi_0)\left[\tanh ( { x_{0}}+X_{{0}}) -\sqrt {1-{\omega}^{2}}\right]^{3}{{e}^{3\,\sqrt {1-{\omega}^{2}}({ x_{0}}+X_{{0}}) }},\\
L_{2}&=&2i\omega\,{ D_{2}B}\left[\tanh ( { x_{0}}-X_{{0}}) -\sqrt {1-{\omega}^{2}}\right]{{e}^{\sqrt {1-{\omega}^{2}}({ x_{0}}-X_{{0}}) }} \\&&-B|B|^2\left( \tilde{E}_{{0}} ( X_{{0}}) +\tilde{E}_{{2}} ( X_{{0}}) \right)\sin(\phi_0-\kappa)\left[\tanh ( { x_{0}}-X_{{0}}) -\sqrt {1-{\omega}^{2}}\right]{{e}^{\sqrt {1-{\omega}^{2}}({ x_{0}}-X_{{0}}) }} \\&&
-\frac{1}{2}{B|B|^2}\cos(\phi_0-\kappa)\left[\tanh ( { x_{0}}-X_{{0}}) -\sqrt {1-{\omega}^{2}}\right]^{3}{{e}^{3\,\sqrt {1-{\omega}^{2}}({ x_{0}}-X_{{0}}) }}.
\end{eqnarray*}
 The solvability condition of the equation will give us
\begin{eqnarray*}
D_{2}B=m_{1} B|B|^{2},
\label{eq13}
\end{eqnarray*}
where
\begin{eqnarray*}
m_{1}&=&-\frac{\Upsilon\, i }{\Psi_{1}(\omega)},\\
\Upsilon &=&\int_{-\infty }^{0}\!f_{{1}}( X_{{0}})dX_{0}+\int _{0}^{\infty }\!f_{{2}}( X_{{0}})dX_{0}+\frac{1}{2}\int_{-\infty }^{0}\!f_{{3}}( X_{{0}})dX_{0}+\frac{1}{2} \int _{0}^{\infty }\!f_{{4}}( X_{{0}})dX_{0},\\
\Psi_{1}(\omega)&=&\frac{2\,\omega\,{{e}^{2\,\sqrt {1-{\omega}^{2}}x_{{0}}}}\Big(2\,\sqrt {1-{\omega}^{2}}+2-{\omega}^{2} -{{e}^{2\,x_{{0}}}}\left( 2\sqrt {1-{\omega}^{2}}+\,{\omega}^{2}-2\right)\Big)}{ \sqrt{1-{\omega}^{2}}(1+ {{e}^{2x_{{0}}}})},\\
f_{{1}}(X_{0})&=&-2\left( E_{0}( X_{{0}}) +E_{2}( X_{{0}})  \right) \sech( x_{0}+X_{{0}})\tanh({ x_{0}}+X_{{0}}) \\&&
{[\tanh \left( {\it x_{0}}+X_{{0}} \right) -\sqrt {1-{\omega}^{2}}]}^{2}{{e}^{2\,\sqrt {1-{\omega}^{2}} \left( {\it x_{0}}+X_{{0}} \right) }},\\
f_{{2}}(X_{0})&=&2\left(\tilde{E_{0}}( X_{{0}}) +\tilde{E_{2}}( X_{{0}})  \right) \sech({ x_{0}}-X_{{0}})\tanh({ x_{0}}-X_{{0}})\\&&
{\Big[\tanh (x_{0}-X_{{0}}) -\sqrt {1-{\omega}^{2}}\Big]}^{2}{{e}^{2\,\sqrt {1-{\omega}^{2}} \left(x_{0}-X_{{0}} \right) }},\\
f_{{3}}(X_{0})&=&{\Big[\tanh \left( {\it x_{0}}+X_{{0}} \right) -\sqrt {1-{\omega}^{2}}\Big]}^{4}(1-2\,\rm{sech\,}^{2}({ x_{0}}+X_{{0}})){{e}^{4\,\sqrt {1-{\omega}^{2}} \left( {\it x_{0}}+X_{{0}} \right) }},\\
f_{{4}}(X_{0})&=&{\Big[\tanh \left( {\it x_{0}}-X_{{0}} \right) -\sqrt {1-{\omega}^{2}}\Big]}^{4}(1-2\,\rm{sech\,}^{2}({ x_{0}}-X_{{0}})){{e}^{4\,\sqrt {1-{\omega}^{2}} \left( {\it x_{0}}-X_{{0}} \right) }}.
\end{eqnarray*}
We will not continue the perturbation expansion to higher orders as we have obtained the leading order behavior of the wobbling amplitude.

Using the chain-rule  and writing $b=\epsilon B$, we obtain
\begin{eqnarray}
\frac{\partial b}{\partial{t}}=m_{1}b|b|^{2}.
\label{a12}
\end{eqnarray}
 It can be derived that
\begin{eqnarray*}
\frac{\partial |b|^{2}}{\partial t}=2\,{\rm Re}( m_{1})|b|^{4},
\end{eqnarray*}
with the solution given by
\begin{eqnarray}
|b|=\sqrt{\frac{|b(0)|^{2}}{1-2\,{\rm Re}( m_{1})|b(0)|^{2}t}},
\label{a9}
\end{eqnarray}
and the initial amplitude $b(0).$  It can be clearly seen that the oscillation amplitude of the breathing mode decreases in time with order $\mathcal{O}({t^{-1/2}})$.

\begin{remark}
Similarly to Remark \ref{r1}, the $\mathcal{O}({t^{-1/2}})$ amplitude decay is caused by the assumption (\ref{2w1}). One therefore can introduce a similar proposition as before. This then leads us to the following conjecture.
\begin{proposition}
If $n\geq2$ is an integer such that
\[
1/(n-1)^2>\omega^2>1/n^2,
\]
then the decay rate of the breathing mode oscillation in $0-\kappa$ Josephson junctions is of order $\mathcal{O}(t^{-1/n})$.
\end{proposition}
\end{remark}

\section{ Driven breathing modes in a $0-\kappa $ junction}
We now consider breathing mode oscillations in  $0-\kappa $ junctions in the presence of external driving with frequency near the natural breathing frequency of the mode, i.e.\ (\ref{eq1}) and (\ref{th2}) with $h\neq0$ and $\Omega=\omega(1+\rho)$. Taking the same scaling as in the case of driven $0-\pi-0$ junction, we obtain (\ref{a4}). Yet, here we assume that the driving amplitude and frequency are small, i.e.\
\begin{equation}
h=\epsilon^{2} H, \,\rho=\epsilon^{2} R,
\end{equation}
with $H, R \sim \mathcal{O}(1)$.

Performing the same perturbation expansion as before, up to  $\mathcal {O(\epsilon)}$ we obtain the same equations as in the undriven case, which we omit for brevity.

The equation at $ \mathcal{O}(\epsilon^{2})$ in the perturbation expansion is
\begin{eqnarray}
\partial_0^2\phi_2-D_0^2\phi_2-\cos(\theta+\phi_0)\phi_2&=&2\left(D_0D_1-\partial_0\partial_1\right)\phi_1-\frac{\phi_1^2}{2} \sin(\theta+\phi_0)\\&&+\frac12H\left(e^{i\omega \tau}+c.c.\right)\nonumber.
\end{eqnarray}
Again, one can write the solution $\phi_2$ as a combination of solutions with harmonics present in the right hand side. In this case, the first harmonic component is different from the undriven case due to the driving, yielding the solvability condition
\begin{eqnarray}
 D_{1}B=mHi,
 \label{eq15}
\end{eqnarray}
 where
\begin{eqnarray*}
m&=&\frac{\eta(x_{0},\omega)}{2\Psi_{1}(\omega)},\\
\eta(x_{0},\omega)&=&\int _{-\infty }^{0}\!{{e}^{\sqrt {1-{\omega}^{2}}\left({\it x_{0}}+X_{{0}}\right)}}\Big[\tanh ({\it x_{0}}+X_{{0}}) -\sqrt {1-{\omega}^{2}}\Big]{dX_{{0}}} \\&&+\int _{0}^{\infty }\!{{e}^{\sqrt {1-{\omega}^{2}}\left({\it x_{0}}-X_{{0}}\right) }} \Big[ \tanh \left( {\it x_{0}}-X_{{0}}\right) -\sqrt {1-{\omega}^{2}}\big] {dX_{{0}}}.
\end{eqnarray*}
The solution for the first harmonic is then
\begin{eqnarray}
\phi^{(1)}_{2}(X_{0},T_{0})=H\left\{
\begin{array}{cc}
g(X_{0})+ n(X_{0}),
& X_{0}<0, \\
\tilde{g}(X_{0})+\tilde{n}(X_{0}),
 & X_{0}>0,
\end{array}
\right.
\end{eqnarray}
where
\begin{eqnarray}
g(X_{0})&=&\frac{\left(2\,\sqrt {1-{\omega}^{2}} +2-{\omega}^{2}-{\omega}^{2}{{e}^{2\,( x_{{0}}+X_{0} )  }} \right){{e}^{\sqrt {1-{\omega}^{2}}( x_{{0}}+X_{0} ) }}C_{g1}}{1+{{e}^{2(\,x_{{0}}+X_{0}}})}\label{v11}\\&&
+\frac{\eta(x_{0},X_{0})}{2\Psi_{1}(\omega)}g_{1}(X_{0}),\nonumber\\
\tilde{g}(X_{0})&=&\frac{\left(2\,\sqrt{1-{\omega}^{2}}-2+{\omega}^{2}+{\omega}^{2}{{e}^{-2\,( x_{{0}}-X_{0}) }} \right){{e}^{\sqrt {1-{\omega}^{2}}( x_{{0}}-X_{0} ) }}C_{g2}}{1+{{e}^{-2(\,x_{{0}}-X_{0}}})}\label{v12} \\&&-\frac{\eta(x_{0},X_{0}) }{2\Psi_{1}( \omega)} \tilde{g}_{1}(X_{0}),\nonumber \\
n({X_{0}})&=&\frac{\left(2\,\sqrt {1-{\omega}^{2}}+2-{\omega}^{2} -{\omega}^{2}{{e}^{2\,( x_{{0}} +X_{0}) }}\right){{e}^{\sqrt {1-{\omega}^{2}}( x_{{0}}+X_{0} ) }}\tilde{C_{h1}}}{1+{{e}^{2(\,x_{{0}}+X_{0})}}}\label{v13}\\&&+\frac{\left({\omega}^{2}{{e}^{ 2\,( x_{{0}}+X_{0} )}}- 2\,\sqrt {1-{\omega}^{2}}-2+{\omega}^{2}\right){{e}^{\sqrt {1-{\omega}^{2}}( x_{{0}}+X_{0} ) }}A_{1}(X_{0})}{4{\omega}^{4}\sqrt {1-{\omega}^{2}}(1+{{e}^{2 ( x_{{0}} +X_{0})}})}\nonumber \\&&
+\frac{\left({\omega}^{2}{{e}^{2\,( x_{{0}}+X_{0} )  }}+ 2\,\sqrt {1-{\omega}^{2}}-2+{\omega}^{2}\right){{e}^{-\sqrt {1-{\omega}^{2}}( x_{{0}}+X_{0} ) }}A_{2}(X_{0})}{4{\omega}^{4}\sqrt {1-{\omega}^{2}}(1+{{e}^{2(\,x_{{0}}+X_{0})}})},\nonumber\\
\tilde{n}(X_{0})&=&\frac{\left({\omega}^{2}{{e}^{-2\,(x_{{0}}-X_{0} )  }}+ 2\,\sqrt {1-{\omega}^{2}}-2 +{\omega}^{2} \right){{e}^{\sqrt {1-{\omega}^{2}}(x_{{0}}-X_{0} ) })}\tilde{C_{h2}}}{1+{{e}^{-2(\,x_{{0}}-X_{0})}}}\label{v14}\\&&+\frac{\left({\omega}^{2}{{e}^{-2\,(x_{{0}}-X_{0} )}}- 2\,\sqrt {1-{\omega}^{2}}+2-{\omega}^{2}\right){{e}^{-\sqrt {1-{\omega}^{2}}(x_{{0}}-X_{0} ) }}A_{3}(X_{0})}{4{\omega}^{4}\sqrt {1-{\omega}^{2}}(1+{{e}^{-2(\,x_{{0}}-X_{0})}})}\nonumber\\&&
+\frac{({{e}^{-2\,( x_{{0}}-X_{0}) }}{\omega}^{2}+ 2\,\sqrt {1-{\omega}^{2}}- 2+{\omega}^{2} ){{e}^{\sqrt {1-{\omega}^{2}}( x_{{0}}-X_{0} ) }}A_{4}(X_{0})}{4{\omega}^{4} \sqrt {1-{\omega}^{2}}(1+{{e}^{-2(\,x_{{0}}-X_{0})}})}.\nonumber
\end{eqnarray}
Here, ${g}_{1}(X_{0})$, $\tilde{g}_{1}(X_{0})$, and $A_j$, $j=1,\dots,4,$ are given in (\ref{g1})--(\ref{A4}). $C_{g1}$, $C_{g2}$, $\tilde{C_{h1}}$ and $\tilde{C_{h2}}$ are constant of integration chosen to satisfy continuity conditions. The other harmonics are the same as in the undamped, undriven case.


Equating the terms at $\mathcal{O}(\epsilon^3)$ gives the equation
\begin{eqnarray*}
\partial_0^2\phi_3-D_0^2\phi_3-\cos(\theta+\phi_0)\phi_3&=&2(D_0D_2 -\partial_0\partial_2)\phi_1+(D_1^2-\partial_1^2)\phi_1  +2(D_0D_1 -\partial_0\partial_1)\phi_2 \\&& +2RD_{0}^{2}\phi_{1} -\phi_1\phi_2\sin(\theta+\phi_0)-\frac{\phi_1^3}{6}\cos(\theta+\phi_0)\nonumber.
\end{eqnarray*}
The solvability condition for the first harmonic of the above equation gives
\begin{eqnarray}
D_{2}B=m_{1} B|B|^{2}-\omega B R\,i.
\label{eq16}
\end{eqnarray}
Equations (\ref{eq15}) and (\ref{eq16}) are the leading order equations governing the oscillation amplitude of the breathing mode. Combining both equations gives us
\begin{eqnarray}
\frac{\omega}{\Omega}\frac{\partial b}{\partial{t}}=mhi+m_{1}b|b|^{2}-i\omega b\rho.
\label{a10}
\end{eqnarray}
Similarly to (\ref{a8}), one can also expect to obtain the result that a nonzero external drive amplitude induces breathing mode oscillation.

\section{Numerical calculations}

\begin{figure}[tbhp!]
 \begin{center}
   \includegraphics[width=0.6\textwidth,angle=0]{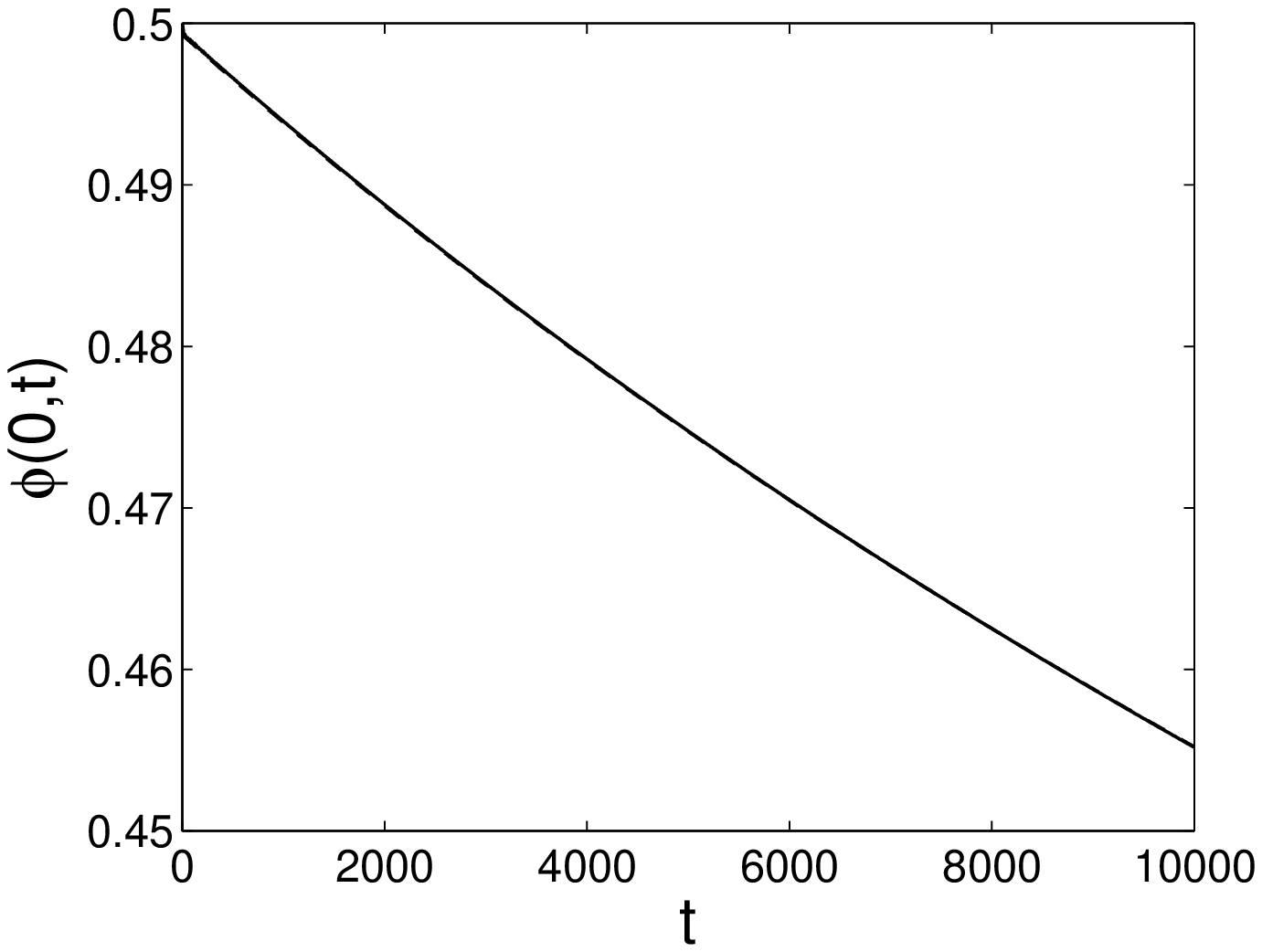}
  \includegraphics[width=0.6\textwidth,angle=0]{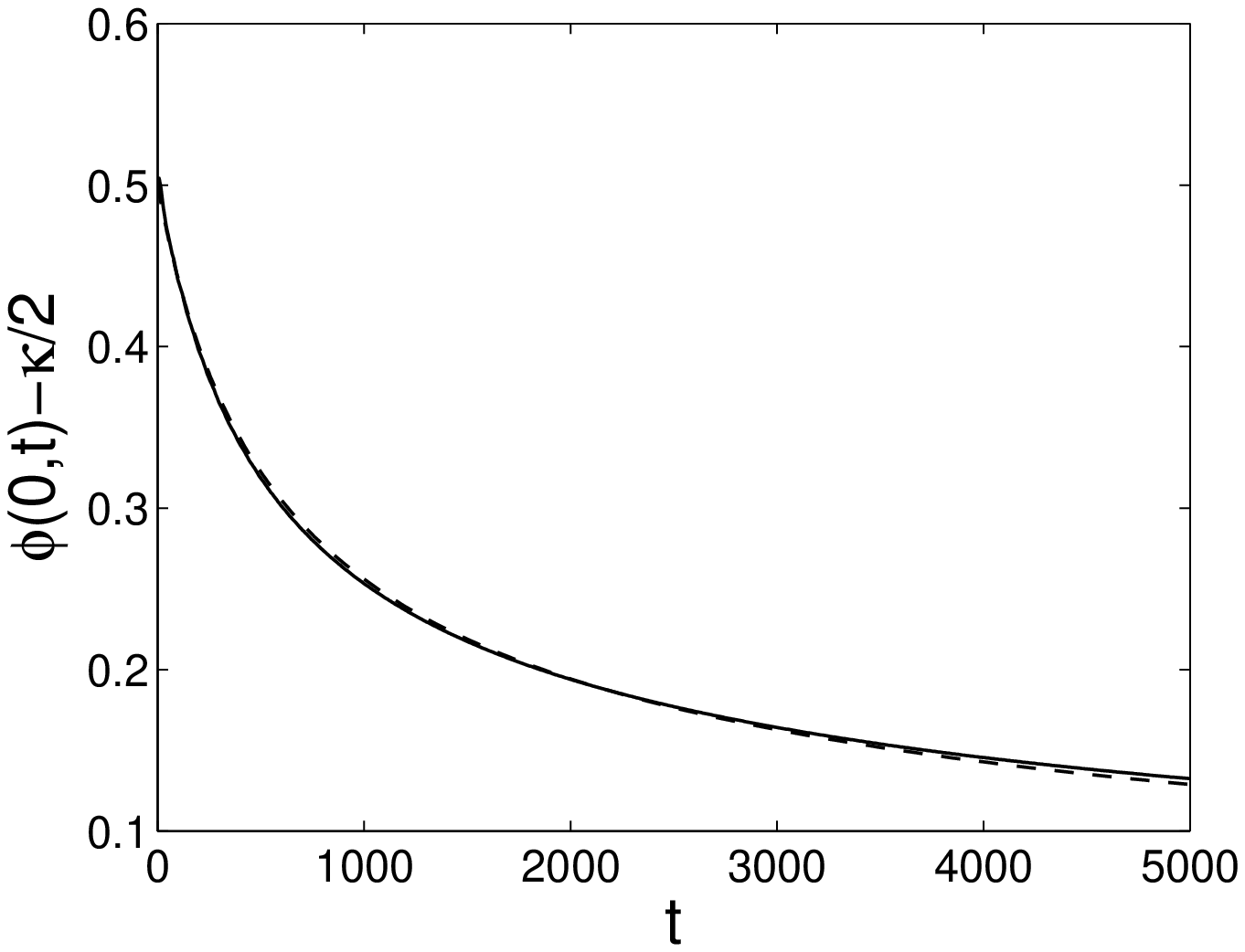}
 \end{center}
  \caption{ Oscillation amplitude of the breathing mode in a $0-\pi-0$ (top panel) and $0-\kappa$ (bottom panel) junction. The solid curves are from the original governing equation (\ref{eq1}), clearly indicating the decay of the oscillation. Analytical approximations (\ref{a3}) for the top panel and (\ref{a9}) for the bottom panel are shown in dashed lines (see the text).}
  \label{fig2}
\end{figure}

To check our analytical results obtained in the above sections, we have numerically solved the governing equation (\ref{eq1}) with $\theta(x)$  given by (\ref{th1}) or (\ref{th2}). We discretize the Laplacian operator using a central difference and integrate the resulting system of differential equations using a fourth-order Runge-Kutta method, with a spatial and temporal discretization $\Delta x=0.02$ and $\Delta t=0.004$, respectively. The computational domain is $x\in(-L,L)$, with $L=50$. At the boundaries, we use a periodic boundary condition. To model an infinitely long junction, we apply an increasing damping at the boundaries to reduce reflected continuous wave incoming from the boundaries. In all the results presented herein, we use the damping coefficient
\begin{eqnarray}
\alpha =
\left\{
\begin{array}{ll}
\left(|x|-L+x_\alpha)\right)/x_\alpha,\,|x|>(L-x_\alpha),\\
0,\,|x|<(L-x_\alpha),
\end{array}
\right.
\end{eqnarray}
i.e.\ $\alpha$ increases linearly from $\alpha=0$ at $x=\pm(L-x_\alpha)$ to $\alpha=1$ at $x=\pm L$. We have taken $x_\alpha=20.$ To ensure that the numerical results are not influenced by the choice of the parameter values above, we have taken different values as well as different boundary conditions and damping, where quantitatively we obtained relatively the same results.

In this section, for the $0-\pi-0$ junction we fix the facet length $a=0.4$, which implies that $\omega\approx0.73825$, and for the $0-\kappa$ junction we set $\kappa=\pi$, which implies that $x_{0}\approx-0.8814$ and $\omega\approx0.8995$. For the choice of parameters above, we obtain  the coefficients in the analytically obtained approximations (\ref{a2}), (\ref{a8}), (\ref{a12}), and (\ref{a10}) as
\begin{equation}\begin{array}{lll}
&k_{1}=0.0433,\,\quad &k_{2}=-0.00324-0.01400\,i\\
&l_{1}=0.6068,\,\quad &l_{2}=-0.10027,\quad l_3=0.05934,\\
&m=- 0.6236,\,\quad &m_{1}=-0.0182-0.0809\,i.
\end{array}
\nonumber
\end{equation}

First, we consider the undriven case, $h=0$. With the initial condition (\ref{init}) and $B_0$ particularly taken to be
\begin{equation}
B_0=\frac{0.5}{\Phi_1(0,0)},
\label{B0}
\end{equation}
where $\Phi_1(x,t)$ is given by (\ref{p11}) for $0-\pi-0$ junctions and (\ref{p12}) for $0-\kappa$ junctions, we record the \emph{envelope} of the oscillation amplitude $\phi(0,t)$ from the governing equation (\ref{eq1}). In Fig.\ \ref{fig2}, we plot in solid lines $\phi(0,t)$ of the $0-\pi-0$ and the $0-\kappa$ junction in the top and bottom panel, respectively.

From Fig.\ \ref{fig2}, one can see that the oscillation amplitude decreases in time. The mode experiences damping. The damping is intrinsically present because the breathing mode emits radiation due to higher harmonics excitations with frequency in the dispersion relation.

\begin{figure}[tbp!]
  \begin{center}
    \includegraphics[width=0.6\textwidth,angle=0]{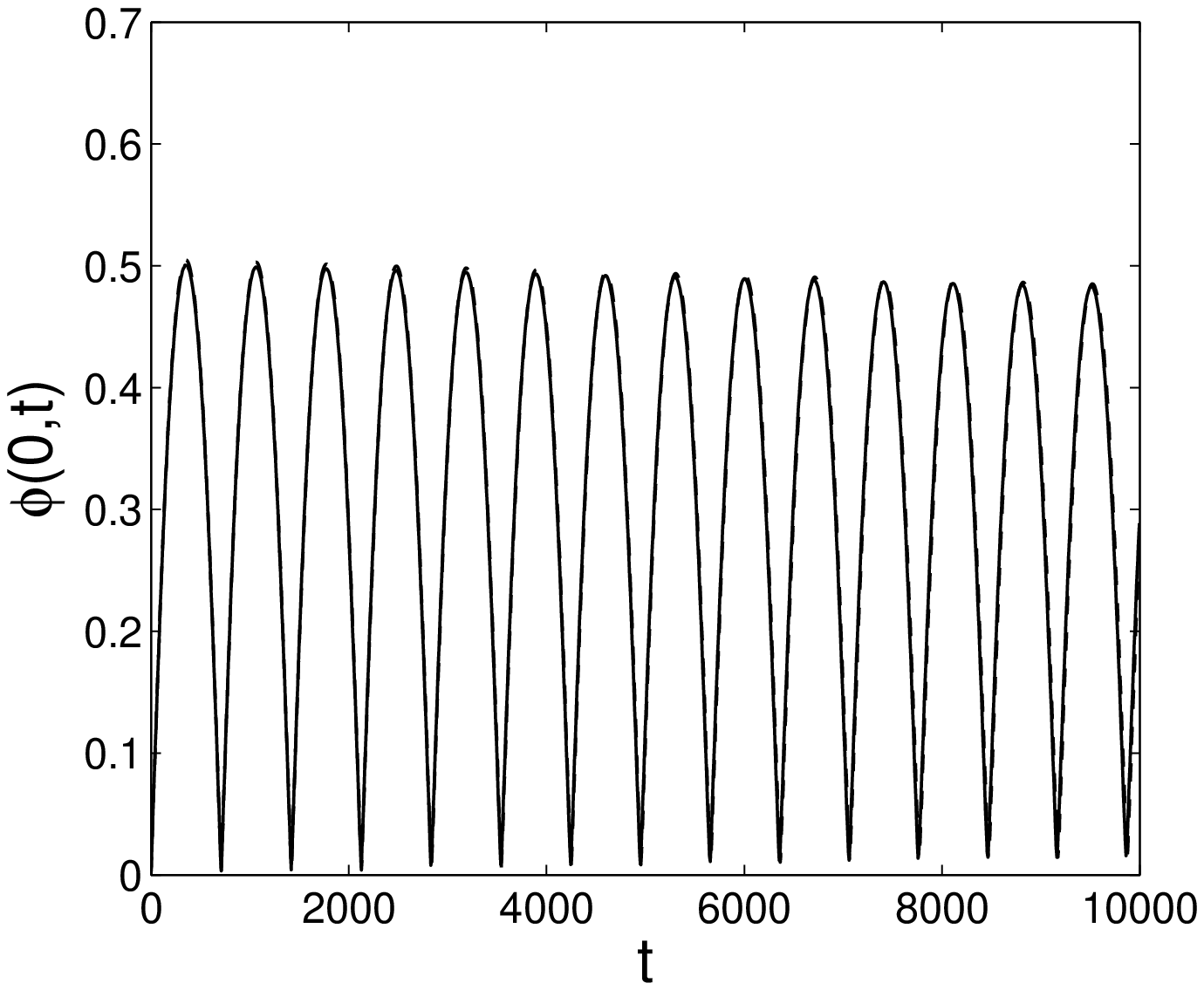}
    \includegraphics[width=0.6\textwidth,angle=0]{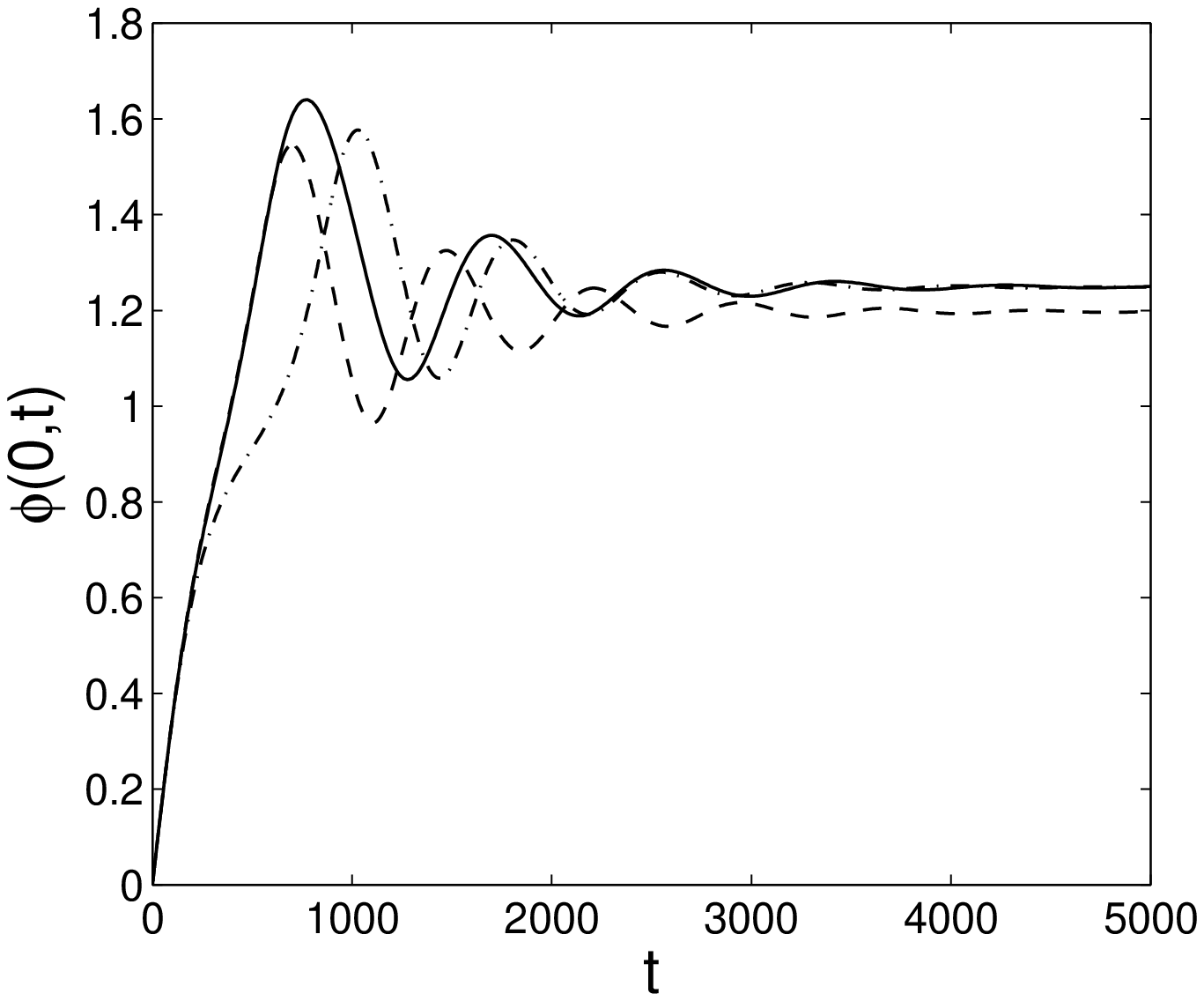}
    \includegraphics[width=0.6\textwidth,angle=0]{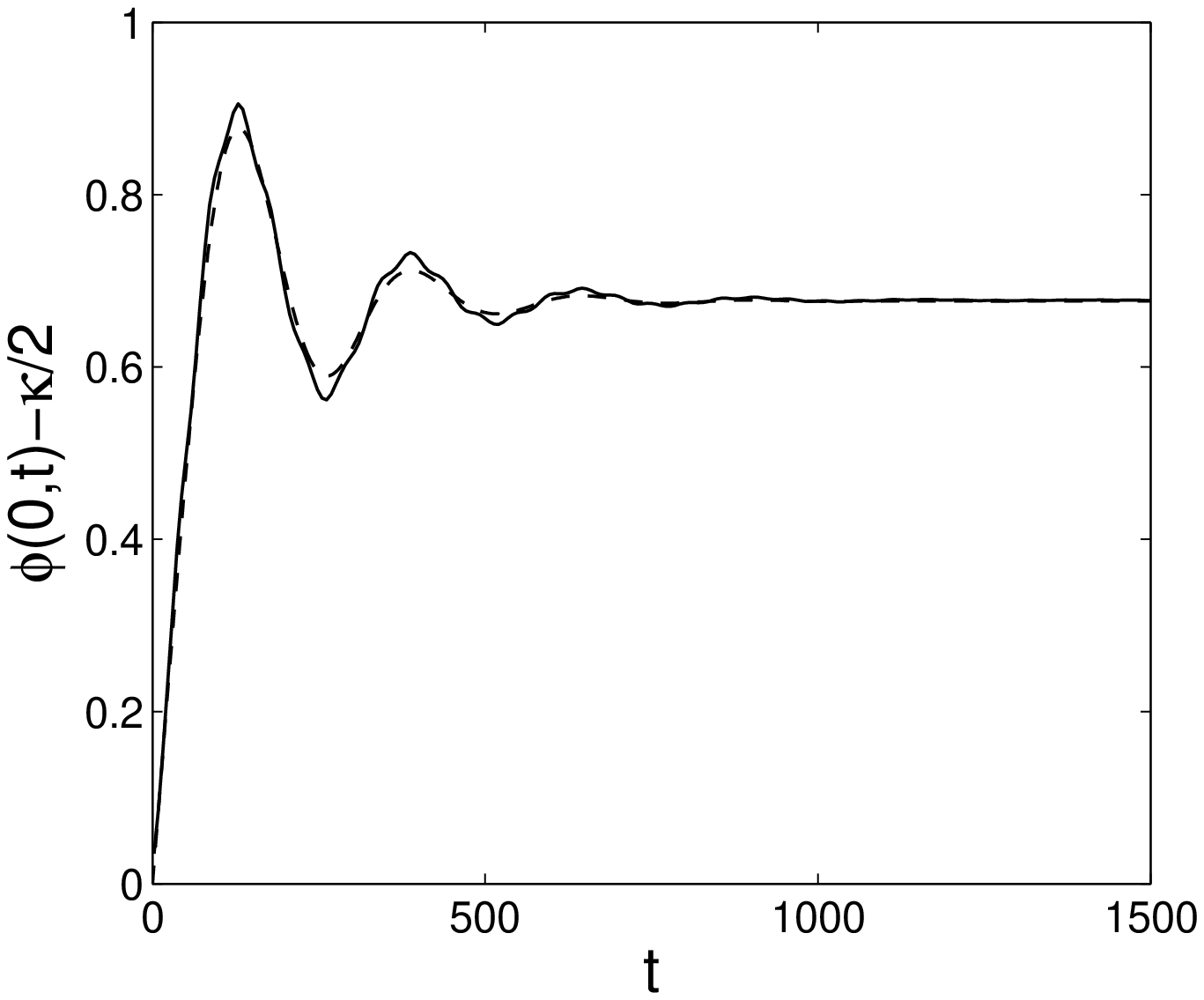}
  \end{center}
  \caption{The same as in Fig.\ \ref{fig2}, but for nonzero driving amplitude. Top and middle panel correspond to driven $0-\pi-0$ junctions with $h=0.002$ and $h=0.003$, respectively. Bottom panel corresponds to driven $0-\kappa$ junctions with $h=0.01$.}
  \label{fig3}
\end{figure}

It is then instructive to compare the numerical results with our analytical calculations. With the initial condition
\[
b(0)=B_0/F,
\]
the analytical approximation is then given by $F|b(t)|$, where $|b(t)|$ is given by (\ref{a3}) and (\ref{a9}) for the $0-\pi-0$ and $0-\kappa$ Josephson junctions, respectively.

In general, the factor $F$ is simply $F=2\Phi_1(0,0),$ i.e.\ $F=2$ for the $0-\pi-0$ case and $F\approx1.56$ for the $0-\kappa$ case. Yet, by treating $F$ as a fitting parameter we observed that the best fit is not given by the aforementioned values. For the initial condition (\ref{B0}), we found that an optimum fit is respectively provided by $F=2.06$ and $F=1.8$. Shown in dashed line in Fig.\ \ref{fig2} is our approximation, where one can see a good agreement between the numerically obtained oscillation and its approximations. In the top panel, the approximation coincides with the numerical result.

Next, we consider the case of driven Josephson junctions, i.e.\ (\ref{eq1}) with $h\neq0$. In this case, the initial condition to the governing equation (\ref{eq1}) is (\ref{init}) with $B_0$ particularly chosen to be
\[B_0=0.\]
Taking specifically $\Omega=\omega$, respectively we present the amplitude of the oscillatory mode $\phi(0,t)$ of $0-\pi-0$ junctions with $h=0.002$ and $h=0.003$ in the top and middle panel of Fig.\ \ref{fig3}.  These are the typical dynamics of the oscillation amplitude of the breathing mode, where for the first case the envelope oscillates periodically in a long time scale and for the second case the amplitude tends to a constant. 

To compare it with the asymptotic analysis,  we have solved the amplitude equations (\ref{a8}) and (\ref{a10}) numerically using a fourth-order Runge-Kutta method with a relatively fine time discretization parameter, as exact analytical solutions are not available. The analytical approximation is again given by $F|b(t)|$, where $F$ in this case is taken to be exactly $F=2\Phi_1(0,0)$. It is important to note that ideally $\rho=0$ as the driving frequency was taken to be the same as the internal frequency of the infinitely long continuous Josephson junctions. Yet, one needs to note that to simulate the governing equation numerically, the equation is discretized and solved on a finite interval, which implies that the system's internal frequency is likely different from the original equation. Therefore, $\rho$ may not be necessarily zero.

Treating $\rho$ as a fitting parameter, we are able to find a good agreement between the numerics and the approximations for $\rho\approx0$. Shown in the top panel of Fig.\ \ref{fig3} in dashed line is the approximation (\ref{a8}) using $\rho=0.00607$, where one can see that our approximation is in a good agreement as it is rather undistinguishable from the numerical result. In the middle panel of the same figure in dashed and dash-dotted line are the approximations for the driving amplitude $h=0.003$ with $\rho=0.006$ and $\rho=0.00665$, respectively. The two values of $\rho$ give a good approximation in different time intervals. It is surprising to see that the amplitude equation is still able to quantitatively capture the numerical result considering the large amplitude produced by the forcing, which is rather beyond the smallness assumption of the oscillation amplitude.

In the bottom panel of Fig.\ \ref{fig3}, we plot the  amplitude of the breathing mode in the $0-\kappa$ junction case with $h=0.01$, where one can see that the envelope of the oscillation amplitude tends to a constant. The dashed curve depicts our approximation from (\ref{a10}) with $\rho=-0.0015$, where a good agreement is obtained. 

Considering the panels in Fig.\ \ref{fig3}, we  observe that the mode in the two junction types does not oscillate with an unbounded or growing amplitude. After a while, there is a balance of energy input into the breathing mode due to the external drive and the radiative damping. The regular oscillation of the mode in the top panel indicates that the junction voltage vanishes, even when the driving frequency is the same as the system's eigenfrequency. 
This  raises the question as to whether the breathing mode of a junction with a phase-shift can be excited further by increasing the driving amplitude to switch the junction to a nonzero voltage. To answer this question, we have solely used numerical simulations of (\ref{eq1}) as it is beyond our perturbation analysis.

\begin{figure}[tbhp!]
  \begin{center}
    \includegraphics[width=0.4\textwidth,angle=0]{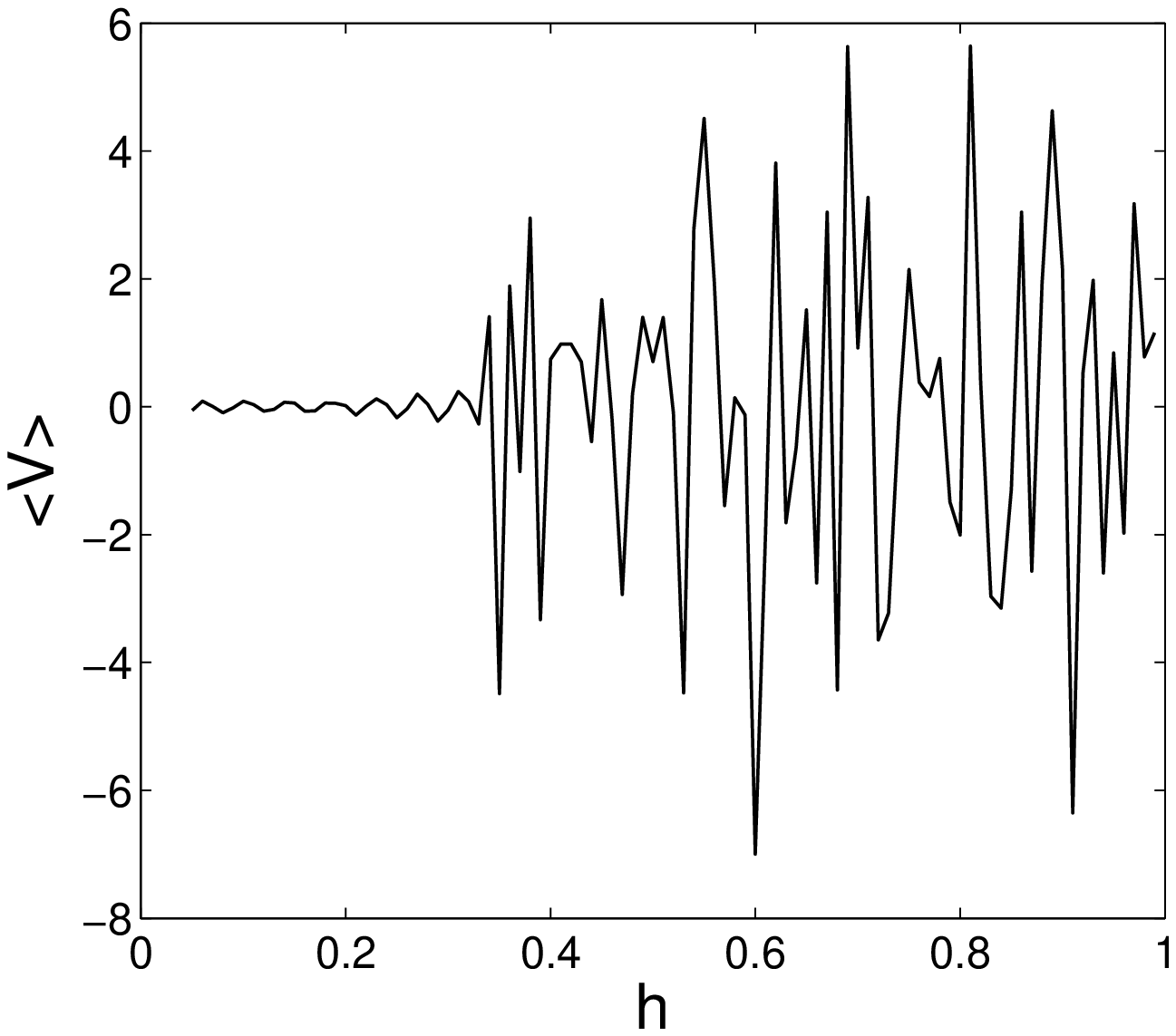}
    \includegraphics[width=0.4\textwidth,angle=0]{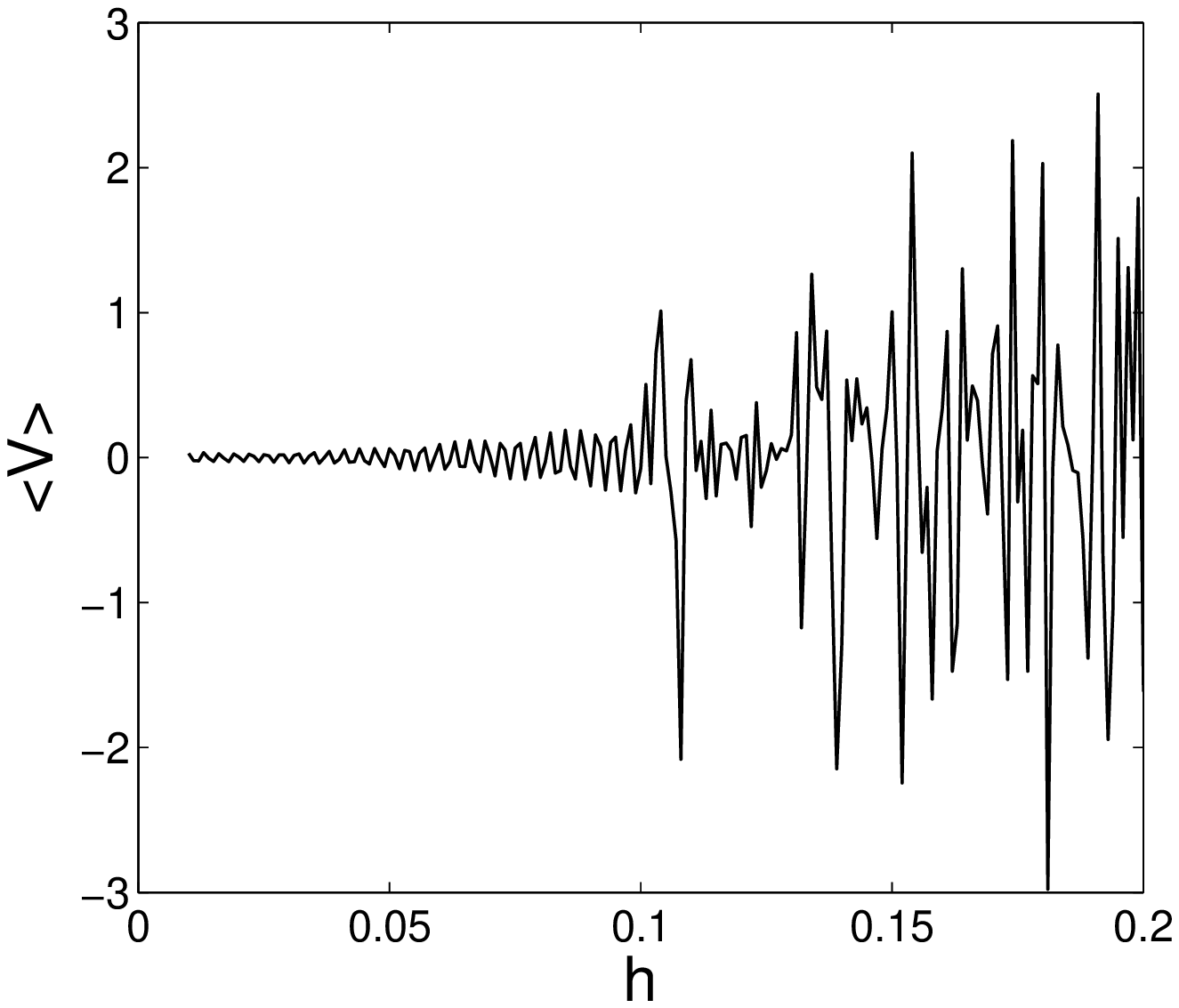}\\
    \includegraphics[width=0.4\textwidth,angle=0]{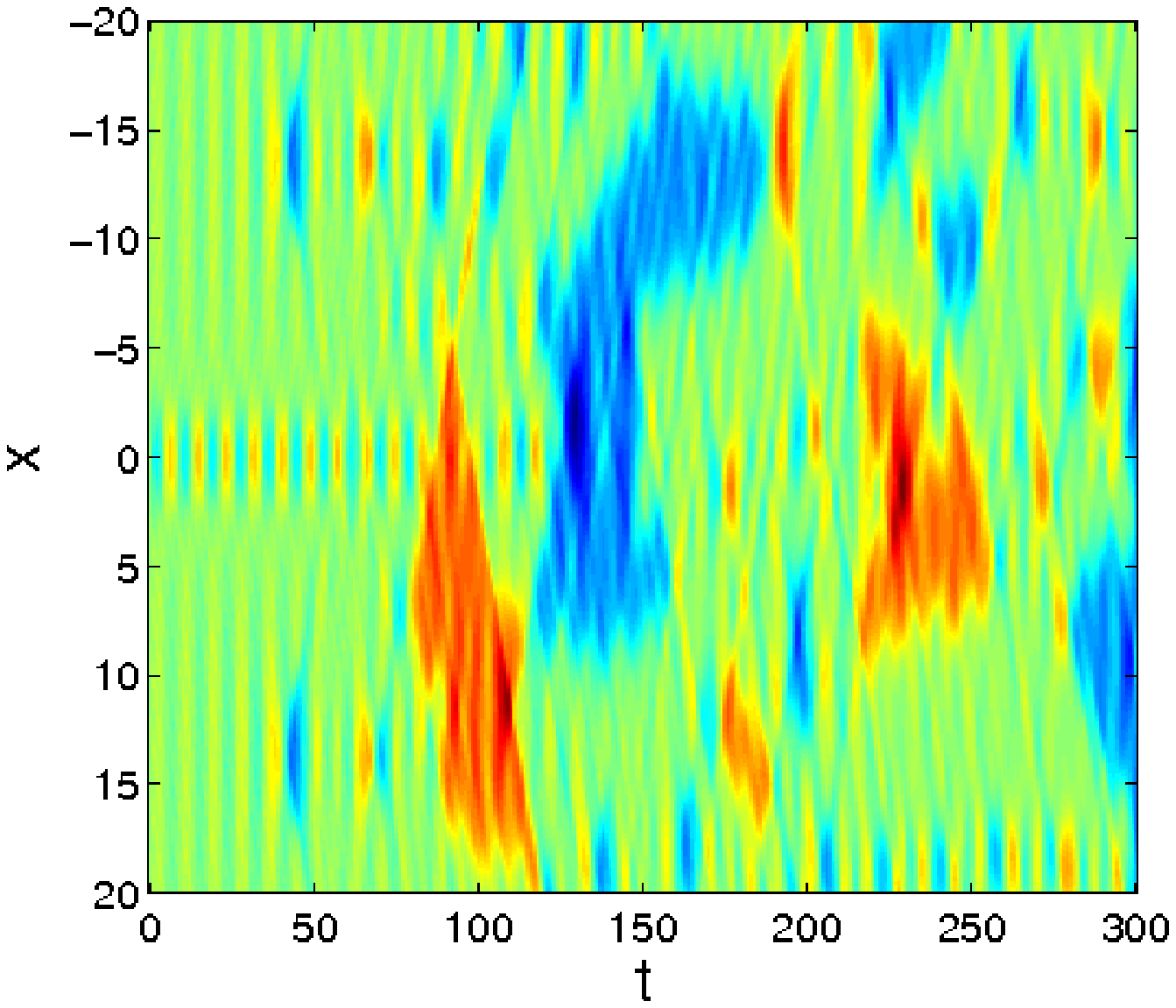}
    \includegraphics[width=0.4\textwidth,angle=0]{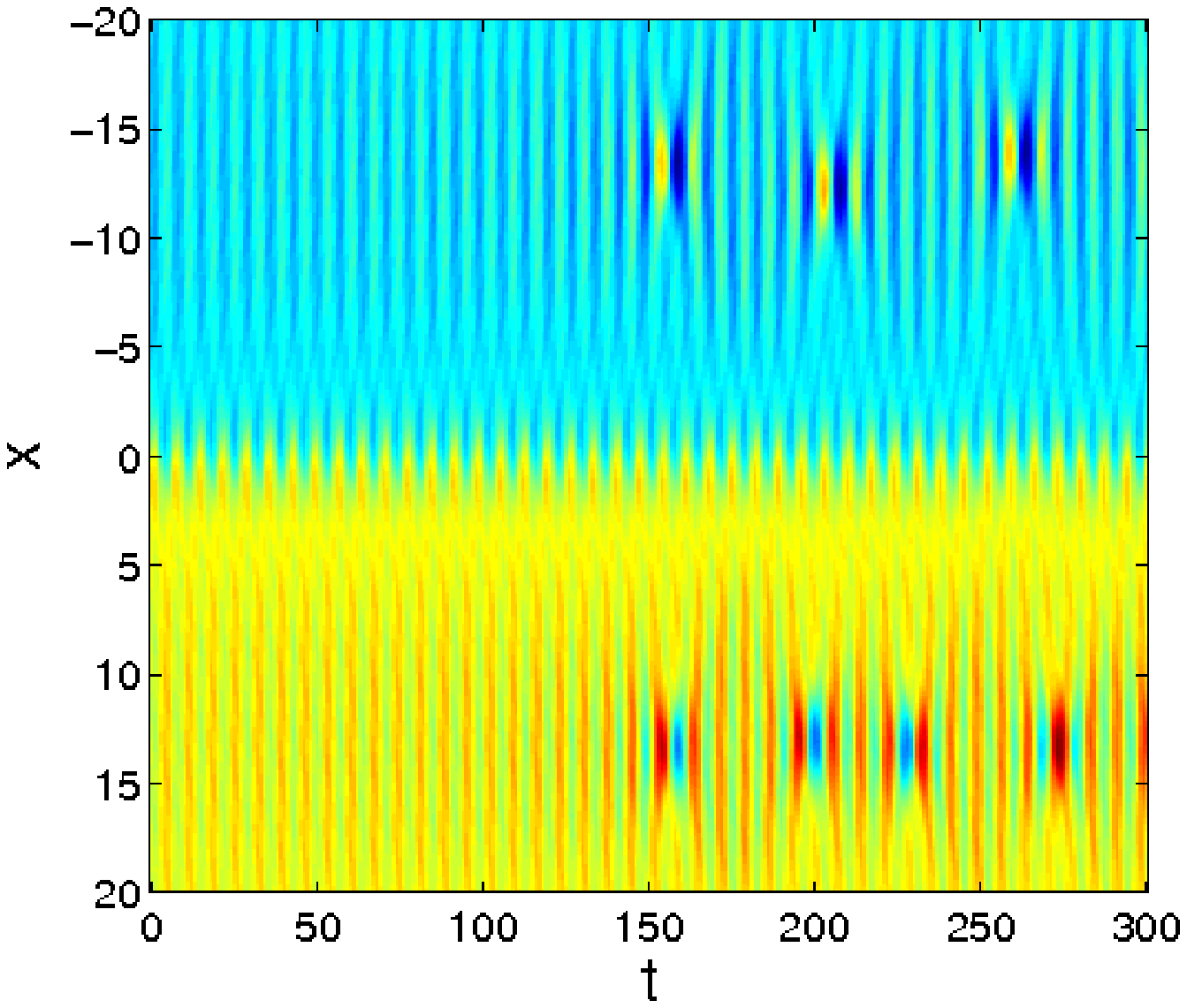}
  \end{center}
  \caption{The average voltage $<V>$ as a function of the driving amplitude $h$ in a $0-\pi-0$ (top left) and $0-\kappa$ (top right) junction, respectively. Bottom panels show the dynamics at the switching point, where the voltage becomes nonzero.}
  \label{fig4}
\end{figure}

In the top left and right panels of Fig.\ \ref{fig4}, we present the average voltage (\ref{V}) with $T=100$ as a function of the external driving amplitude $h$ for the case of $0-\pi-0$ and $0-\kappa$ junctions, respectively. One can clearly see that in both cases, there is a minimum amplitude above which the junction has a large nonzero voltage. For the first and second junction, the critical amplitude is respectively $h\approx0.34$ and $h\approx0.1$. The time dynamics of the transition from the superconducting state $V\approx0$ to a resistive state $V\gg0$ is shown in the bottom panels of the same figure.

Form the panels, it is important to note that apparently the switch from a superconducting to a resistive state is not caused by the breathing mode, but rather because of the continuous wave background emitted by the breathing mode. It shows that the continuous wave becomes modulationally unstable and as typical dynamics breathers are created which later 
interact and destroy the breathing mode. Hence, we  conclude that a breathing mode in these cases cannot be excited to make the junction resistive by applying an external drive, even with a relatively large driving amplitude.

\section{Conclusions}
We have considered a spatially inhomogeneous sine-Gordon equation with a time-periodic drive, modeling a microwave driven long Josephson junction with phase-shifts. Using multiple scale expansions, we have shown that in an infinitely long Josephson junction, an external drive cannot excite the defect mode of a junction, i.e.\ a breathing mode, to switch the junction into a resistive state. For a small drive amplitude, there will be an energy balance between the energy input given by the external drive and the energy output due to  so-called radiative damping experienced by the mode. When the external drive amplitude is large enough, the junction can indeed switch to a resistive state. Yet, this can also be caused by a modulational instability of the continuous wave emitted by the oscillating mode.

Despite the agreement with the experiments obtained herein, our analysis is based on a simplified model. It is then of interest to extend the study to the case of \emph{(dc) driven long, but finite} Josephson junctions with phase-shifts, as experimentally used in \cite{buck07,pfei09}. 
In microwave-driven finite junctions, the boundaries can be a major  external drive (see, e.g., \cite{goldo02,bark04}), which is not present in the study here. A constant (dc) bias current, which is mentioned to play an important role in the measurements reported in \cite{buck07}, is also not included in our current paper, even though the results presented herein should still hold for small enough constant drive. These are currently being studied and will be reported elsewhere. Another open problem that will be studied is the interaction of multiple defect modes \cite{bamb09} in a Josephson junction with phase-shifts. This is experimentally relevant as the so-called zigzag junctions have been successfully fabricated \cite{hilg03}.

\section*{Acknowledgments}
The author wishes to thank the anonymous referees for their constructive comments and suggestions.

\Appendix

\section{Explicit expressions}
\label{AppA}

The functions $k_{3j}$ and $\tilde{k}_{3j}$ in the expression of $\upsilon_j$ (\ref{phi31_0p0}), $j=1,2,3,$ are given by
\begin{eqnarray}
k_{31}&=&2\,\omega\, \tan^{-1} \left( \sqrt{\frac{1-{\omega}^{2}}{1 +{\omega}^{2}}} \right) \sqrt {1+{\omega}^{2}} \left( {{e}^{3\,u+2 \,X_{{0}} \sqrt {1-{\omega}^{2}}}}(1+{\omega}^{2})+3\,{{e}^{u}} X_{{0}}\sqrt {1-{\omega}^{2}} \right)\label{k31}\\
&&+\frac{2\,{{e}^{3\,u+2\,X_{{0}}\sqrt {1-{\omega}^{2}}}}\omega\, (1+\,{\omega}^{2})^{2}}{\sqrt{1-\omega^{2}}}-{{e}^{u}}\omega\,X_{{0}} \left( 2\,{\omega}^{6}-3 +2\,{\omega}^{2} +7\,{\omega}^{4} \right),\nonumber\\
\tilde{k}_{31}&=&\frac{{{e}^{u+2\,X_{{0}}\sqrt {1-{\omega}^{2}}}}\omega\,\sqrt {1-{\omega}^{4}} \tan^{-1} \left( \sqrt{\frac{1-{\omega}^{2}}{1 +{\omega}^{2}}} \right) \left( {{e}^{2\,u+2\,X_{{0}}\sqrt {1-{\omega}^{2}}}}(1+{\omega}^{2})+3 \right)} {\sqrt{1-\omega^{2}}} \label{k31t}\\
&& +\frac{{{e}^{3\,u+4\,X_{{0}}\sqrt {1-{\omega}^{2}}}}\omega\, \left( 1+\,{\omega}^{2} \right)^{2}} {\sqrt{1 -\omega^{2}}} -\frac{{{e}^{u+2\,X_{{0}}\sqrt {1-{\omega}^{2}}}}\omega\, \left( 2\,{\omega}^{6}-3+2\,{\omega}^{2}+7\, {\omega}^{4} \right)} {2\sqrt{1-\omega^{2}}},\nonumber\\
k_{32}&=&\frac{ \sin \left(2 \sqrt {1+{\omega}^{2}}X_{{0}} \right)  (2\omega^2+3)(\omega^2+1)^2}{4\sqrt{1+\omega^{2}}} +\frac{X_{{0}}}{2} (2\omega^2+3)(\omega^2+1)^2 \label{k32}\\
&&+\frac{2\,  \cos ^{3}\left( \sqrt {1+{\omega}^{2}}X_{{0}} \right) \sin \left( \sqrt {1+{\omega}^{2}}X_{{0}} \right)  \left( \sqrt {1-{\omega}^{4}} \tan^{-1} \left( \sqrt{\frac{1-{\omega}^{2}}{1 +{\omega}^{2}}} \right)+1+{\omega}^{2} \right)}{\sqrt{1+\omega^{2}}},\nonumber\\
\tilde{k}_{32}&=&\frac{ \cos ^{2}\left( \sqrt {1+{\omega}^{2}}X_{{0}} \right) \left( 6\,\sqrt {1-{\omega}^{4}} \tan^{-1} \left( \sqrt{\frac{1-{\omega}^{2}}{1 +{\omega}^{2}}} \right) -7\,{\omega}^{4}-2\,{\omega}^{2}+3-2\,{\omega}^{6} \right)}{2\sqrt{1+\omega^{2}}}\label{k32t}\\
&&- \frac{2\,  \cos^{4} \left( \sqrt {1+{\omega}^{2}}X_{{0}} \right) \left( \sqrt {1-{\omega}^{4}} \tan^{-1} \left( \sqrt{\frac{1-{\omega}^{2}}{1 +{\omega}^{2}}} \right) +1+{\omega}^{2} \right) }{\sqrt{1+\omega^{2}}},\nonumber\\
k_{33}&=& \frac{{{e}^{u-2\,X_{{0}}\sqrt {1-{\omega}^{2}}}} \left( 2\,{\omega}^{6}-3+2\,{\omega}^{2}+7\,{\omega}^{4} \right)}{2\sqrt{1- \omega ^{2}}} -\frac{\left( 1+\,{\omega}^{2} \right)^{2} {{e}^{3\,u-4\,X_{{0}}\sqrt {1-{\omega}^{2}}}}}{\sqrt{1-\omega^{2}}} \label{k33}\\
&&-\frac{\left( {{e}^{2\,u}}+{{e}^{2\,u}}{\omega}^{2}+3\,{{e}^{2\,X_{{0}}\sqrt {1-{\omega}^{2}}}} \right) {{e}^{u-4\,X_{{0}}\sqrt {1-{\omega}^{2}}}}\sqrt {1-{\omega}^{4}} \tan^{-1} \left( \sqrt{\frac{1-{\omega}^{2}}{1 +{\omega}^{2}}} \right)}{\sqrt{1-\omega^{2}}},\nonumber\\
\tilde{k}_{33}&=&-(2\,{{e}^{3\,u-2\,X_{{0}}\sqrt {1-{\omega}^{2}}}})\left( 1+{\omega}^{2} \right)\left(\sqrt {1+{\omega}^{2}} \tan^{-1} \left( \sqrt{\frac{1-{\omega}^{2}}{1 +{\omega}^{2}}} \right)-\frac{( 1+{\omega}^{2})}{\sqrt{1-\omega^{2}}}\right) \label{k33t}\\
&&+{{e}^{u}}X_{{0}} \left( 6\,\sqrt {1-{\omega}^{4}} \tan^{-1} \left( \sqrt{\frac{1-{\omega}^{2}}{1 +{\omega}^{2}}} \right)-7 \, {\omega}^{4}-2\,{\omega}^{2}+3-2\,{\omega}^{6} \right).\nonumber
\end{eqnarray}


The functions $E_{j}(X_{0})$ and $\tilde{E_{j}}(X_{0})$ in $\phi_2^{(j)}$ (\ref{phi20_0k})--(\ref{phi22_0k}) ($j=0,2$) are given by
\begin{eqnarray}
&&E_{0}(X_{0})=\frac{e^{(X_{0}+x_{0})}C_{01}}{1+e^{2(X_{0}+x_{0})}}-\frac{2\, ( 2+\sqrt {1-{\omega}^{2}}) {{e}^{( X_{{0}}+{\it x_{0}})  ( 2\,\sqrt {1-{\omega}^{2}}+3 ) }}}{\sqrt{1-\omega^{2}}(1+e^{2(X_{0}+x_{0})})^{5}}\label{v7}\\&&-\frac{( 1+\sqrt {1-{\omega}^{2}} ) {{e}^{2\,\sqrt {1-{\omega}^{2}} ( X_{{0}}+{\it x_{0}}) +X_{{0}}+{\it x_{0}}}}}{\sqrt{1-\omega^{2}}(1+e^{2(X_{0}+x_{0})})^{5}}\nonumber\\&&
-\frac{2\,( 2-\sqrt {1-{\omega}^{2}}) {{e}^{( X_{{0}}+{\it x_{0}})  \left( 2\,\sqrt {1-{\omega}^{2}}+7 \right) }}+6\,{{e}^{ \left( X_{{0}}+{\it x_{0}} \right)\left( 5+2\,\sqrt {1-{\omega}^{2}} \right) }}}{\sqrt{1-\omega^{2}}(1+e^{2(X_{0}+x_{0})})^{5}}\nonumber\\&&
+\frac{2\,( 1+2\,\sqrt {1-{\omega}^{2}}-{\omega}^{2}) {{e}^{( X_{{0}}+{\it x_{0}})( 2\,\sqrt {1-{\omega}^{2}}+3 ) }}}{(1+e^{2(X_{0}+x_{0})})^{5}},\nonumber\\
&&\tilde{E_{0}}(X_{0})=\frac{e^{(X_{0}-x_{0})}C_{02}}{1+e^{2(X_{0}-x_{0})}}+\frac{2\,{{e}^{( -X_{{0}}+{\it x_{0}} )( 2\,\sqrt {1-{\omega}^{2}} -7) }} ( 2+\sqrt {1-{\omega}^{2}} ) }{\sqrt{1-\omega^{2}}(1+e^{2(X_{0}-x_{0})})^{5}}\label{v8}\\&& -\frac{2\,{{e}^{ \left(-X_{{0}}+{\it x_{0}}\right)\left( 2\,\sqrt {1-{\omega}^{2}} -3\right)}} ( 2-\sqrt {1-{\omega}^{2}})-6\,{{e}^{ \left( -X_{{0}}+{\it x_{0}} \right)  \left( 2\,\sqrt {1-{\omega}^{2}}-5 \right) }}}{\sqrt{1-\omega^{2}}(1+e^{2(X_{0}-x_{0})})^{5}}\nonumber\\&&
+\frac{{{e}^{ \left({\it x_{0}}-X_{{0}} \right)  \left( 2\,\sqrt {1-{\omega}^{2}}-1 \right) }}(1-\sqrt {1-{\omega}^{2}}) +{{e}^{ \left({\it x_{0}}-X_{{0}} \right)  \left( 2\,\sqrt {1-{\omega}^{2}}-9 \right) }} ( 1+\sqrt {1-{\omega}^{2}})}{\sqrt{1-\omega^{2}} (1+e^{2(X_{0}-x_{0})})^{5}},\nonumber\\
&&E_{2}(X_{0})=\frac{(\left( 2\,{\omega}^{2}{{e}^{ 2(X_{0}+x_{0}) }}-(1+\sqrt {1-4\,{\omega}^{2}}- 2\,{\omega}^{2})\right) {{e}^{\sqrt {1-4\,{\omega}^{2}}( X_{{0}}+{\it x_{0}}) }} C_{{21}}}{1+e^{2(X_{0}+x_{0})}}\label{v9}\\
&&-\frac{\left(( 2\,\sqrt {1-{\omega}^{2}}+1) {{e}^{3( X_{{0}}+{\it x_{0}})}}+\, ( 2\,\sqrt {1-{\omega}^{2}}-1 ) {{e}^{7(X_{{0}} +{\it x_{0}} ) }}\right) {{e}^{2\,\sqrt {1-{\omega}^{2}}(X_{{0}} +{\it x_{0}} )  }}}{(1+e^{2(X_{0}+x_{0})})^{5}}\nonumber \\&&
+\frac{-6\,\sqrt {1-{\omega}^{2}}{{e}^{ \left( X_{{0}}+{\it x_{0}} \right)  ( 5+2\,\sqrt {1-{\omega}^{2}}) }}-{{e}^{2\,\sqrt {1-{\omega}^{2}} \left( X_{{0}}+{\it x_{0}} \right) +X_{{0}}+{\it x_{0}}}}( 1+\sqrt {1-{\omega}^{2}}) } {2(1+e^{2(X_{0}+x_{0})})^{5}}\nonumber\\&&
+\frac{( 1-\sqrt {1-{\omega}^{2}}) {{e}^{ \left( X_{{0}}+{\it x_{0}} \right)( 9+2\,\sqrt {1-{\omega}^{2}}) }}}{2((1+e^{2(X_{0}+x_{0})})^{5}},\nonumber\\
&&\tilde{E_{2}}(X_{0})=\frac{ \left( 2\,{\omega}^{2}{{e}^{2\,( X_{{0}}-{\it x_{0}}) }}+2\,{\omega}^{2}+ \sqrt {1-4\,{\omega}^{2}}-1 \right ) {{e}^{-\sqrt {1-4\,{\omega}^{2}}( X_{{0}}-{\it x_{0}}) }}C_{{22}}}{1+e^{2(X_{0}-x_{0})}}\label{v10}\\&&
+\frac{\left({e^{9\,(X_{{0}}-{x_{0}})}}-2\,{e^{3\,(X_{{0}}-{x_{0}})}}-{e^{(X_{{0}}-{ x_{0}}) }}\right){{e}^{-2\,\sqrt {1-{\omega}^{2}}( X_{{0}}-{\it x_{0}}) }}}{2(1+e^{2(X_{0}-x_{0})})^{5}}\nonumber\\&&+\frac{2{ e^{-(X_{{0}}-{x_{0}})(2\sqrt {1-{\omega}^{2}}-7) }}+4{e^{-(X_{{0}}+{ x_{0}})(2\sqrt {1-{\omega}^{2}}-3) }}\sqrt {1-\omega^{2}}}{2((1+e^{2(X_{0}-x_{0})})^{5}}\nonumber\\&&+\frac{\sqrt {1-{\omega}^{2}}( 6\,{{e}^{ -(X_{{0}}-{\it x_{0}} )(2\,\sqrt {1-{\omega}^{2}}-5) }}+{{ e}^{ -( X_{{0}}-{\it x_{0}})(2\,\sqrt{1-{\omega}^{2}}-1) }})}{2(1+e^{2(X_{0}-x_{0})})^{5}}\nonumber \\&& +\frac{ \sqrt{1-{\omega}^{2}}({{e}^{ -(X_{{0}}-{\it x_{0}})(2\,\sqrt {1-{\omega}^{2}}-9) }}+4\,{{e}^{-(X_{{0}}-{\it x_{0}} )( 2\,\sqrt {1-{\omega}^{2}}-7 ) }})}{2(1+e^{2(X_{0}-x_{0})})^{5}}.\nonumber
\end{eqnarray}

The functions ${g}_{1}(X_{0})$, $\tilde{g}_{1}(X_{0})$, and $A_j$, $j=1,\dots,4,$ in (\ref{v11})--(\ref{v14}) are
\begin{eqnarray}
&&\hspace{1cm}g_{1}({X_{0}})=\frac{ ( (  x_{{0}}+X_{0} ) {\omega}^{2}+1) \sqrt {1-{\omega}^{2}}-\frac{1}{2}{\omega}^{2}+1 ) {{e}^{( x_{{0}}+X_{0} ) ( \sqrt {1-{\omega}^{2}}+2 ) }}}{\omega\sqrt {1-{\omega}^{2}}  ( \frac{1}{2}+{{e}^{2(x_{{0}}+X_{0})}}+\frac{1}{2}\,{{e}^{4( x_{{0}} + X_{0} )}})}\label{g1}\\
&&+\frac{(  -2+( x_{{0}}+X_{0} ) {\omega}^{4}-(x_{{0}}+X_{0}-\frac{5}{2}) {\omega}^{2} ) \sqrt {1-{\omega}^{2}}{{e}^{\sqrt {1-{\omega}^{2}}( x_{{0}}+X_{0}) }}} {2\omega\sqrt {1-{\omega}^{2}} ( \omega+1 ) (\omega-1 ) ( \frac{1}{2}+{{e}^{2(x_{{0}}+X_{0})}}+\frac{1}{2}\,{{e}^{4( x_{{0}} + X_{0} )}})}\nonumber\\
&&+\frac{( 2+( -\frac{1}{2}+x_{{0}}+X_{0}) {\omega}^{2} )( \omega^2-1)) {{e}^{\sqrt {1-{\omega}^{2}}( x_{{0}}+X_{0}) }}}{2\omega\sqrt {1-{\omega}^{2}} ( \omega+1 ) (\omega-1 ) ( \frac{1}{2}+{{e}^{2(x_{{0}}+X_{0})}}+\frac{1}{2}\,{{e}^{4( x_{{0}} + X_{0} )}}}\nonumber\\
&&+\frac{{\omega}^{2}( ( x_{{0}}+X_{0}) {\omega}^{2}-x_{{0}}-X_{0}-\frac{1}{2} ) \sqrt {1-{\omega}^{2}}{{e}^{( x_{{0}}+X_{0})( 4+\sqrt {1-{\omega}^{2}}) }}}{2\omega\sqrt {1-{\omega}^{2}} ( \omega+1 ) (\omega-1 ) ( \frac{1}{2}+{{e}^{2(x_{{0}}+X_{0})}}+\frac{1}{2}\,{{e}^{4( x_{{0}} + X_{0} )}})}\nonumber\\
&&+\frac{(( -x_{{0}}-X_{0}-\frac{1}{2}) {\omega}^{2} +x_{{0}}+X_{0}+\frac{1}{2} ) {{e}^{( x_{{0}}+X_{0})( 4+\sqrt {1-{\omega}^{2}}) }}}{2\omega\sqrt {1-{\omega}^{2}} ( \omega+1 ) (\omega-1 ) ( \frac{1}{2}+{{e}^{2(x_{{0}}+X_{0})}}+\frac{1}{2}\,{{e}^{4( x_{{0}} + X_{0} )}})},\nonumber\\
&&\hspace{1cm}\tilde{g_{1}}({X_{0}})=\frac{\, ( ( 1-( x_{{0}}-X_{0} ){\omega}^{2} ) \sqrt {1-{\omega}^{2}}+\frac{1}{2}\,{\omega}^{2}-1 ) {{e}^{( \sqrt {1-{\omega}^{2}}-2)(x_{{0}}-X_{0} ) }}}{\omega\sqrt {1-{\omega}^{2}}  ( \frac{1}{2}+{{e}^{-2(x_{{0}}-X_{0})}}+\frac{1}{2}\,{{e}^{- 4( x_{{0}} - X_{0} )}})}\\
&&+\frac{(( 2+ ( -\frac{1}{2}-x_{{0}}+X_{0} ) {\omega}^{2}) (\omega^2 -1 )+2{\omega}^{4}){{e}^{- \sqrt {1-{\omega}^{2}}( -x_{{0}}+X_{0}) }}}{2\omega\sqrt {1-{\omega}^{2}} ( \omega+1 ) (\omega-1 ) ( \frac{1}{2}+{{e}^{-2(x_{{0}}-X_{0})}}+\frac{1}{2}\,{{e}^{- 4( x_{{0}} - X_{0} )}})}\nonumber\\
&&+\frac{(((X_{0}-x_{{0}} ) {\omega}^{4}+( x_{{0}}- X_{0}+\frac{5}{2} ) {\omega}^{2}) \sqrt{1- {\omega}^{2}}){{e}^{- \sqrt {1-{\omega}^{2}}( -x_{{0}}+X_{0}) }}}{2\omega\sqrt {1-{\omega}^{2}} ( \omega+1 ) (\omega-1 ) ( \frac{1}{2}+{{e}^{-2(x_{{0}}-X_{0})}}+\frac{1}{2}\,{{e}^{- 4( x_{{0}} - X_{0} )}})}\nonumber\\
&&+\frac{ {\omega}^{2}(( -x_{{0}}+X_{0} ) {\omega}^{2}+x_{{0}}-X_{0}-\frac{1}{2} ) \sqrt {1-{\omega}^{2}}{{e}^{- (-x_{{0}} +X_{0} ) ( \sqrt {1-{\omega}^{2}}-4 ) }} }{2\omega\sqrt {1-{\omega}^{2}} ( \omega+1 ) (\omega-1 ) ( \frac{1}{2}+{{e}^{-2(x_{{0}} -X_{0})}} +\frac{1}{2}\,{{e}^{- 4( x_{{0}} - X_{0} )}})}\nonumber\\
&&+\frac{( \omega+1) ( -x_{{0}}+X_{0}+\frac{1}{2} ) ( -1+\omega ) {{e}^{- (-x_{{0}} +X_{0} ) ( \sqrt {1-{\omega}^{2}}-4 ) }}}{2\omega\sqrt {1-{\omega}^{2}} ( \omega+1 ) (\omega-1 ) ( \frac{1}{2}+{{e}^{-2(x_{{0}} -X_{0})}} +\frac{1}{2}\,{{e}^{- 4( x_{{0}} - X_{0} )}})},\nonumber\\
&&A_{1}(X_{0})=\int \!{\frac {( 2\,\sqrt {1-{\omega}^{2}}-2+{\omega}^{2}) {{e}^{-\sqrt {1-{\omega}^{2}} ( x_{{0}}+X_{0} ) }}+{{\omega}^{2}{e}^{-( x_{{0}}+X_{0} ) ( \sqrt {1-{\omega}^{2}}-2) }}}{1+{{e}^{2(\,x_{{0}}+\,X_{0})}}}}{dX_{0}},\\
&&A_{2}(X_{0})=\int \!{\frac {(2\,\sqrt {1-{\omega}^{2}}+2-{\omega}^{2}){{e}^{\sqrt {1-{\omega}^{2}}(x_{{0}}+X_{0})}}-{\omega}^{2}{{e}^{(x_{{0}} +X_{0}) ( \sqrt {1-{\omega}^{2}}+2) }}}{1+{{e}^{2(\,x_{{0}}+\,X_{0})}}}}{d{X_{0}}},\\
&&A_{3}(X_{0})=\int \!{\frac {( 2\,\sqrt {1-{\omega}^{2}}-2+{\omega}^{2} ) {{e}^{\sqrt {1-{\omega}^{2}} (x_{{0}}-X_{0} ) }}+{{\omega}^{2}{e}^{(x_{{0}}-X_{0} )( \sqrt {1-{\omega}^{2}}-2 ) }}}{1+{{e}^{-2(\,x_{{0}}-\,X_{0})}}}}{dX_{0}},\\
&&A_{4}(X_{0})=-\int \!{\frac {{\omega}^{2}{{e}^{-(x_{{0}}-X_{0} )( \sqrt {1-{\omega}^{2}}+2 ) }}+{{e}^{-\sqrt {1-{\omega}^{2}}(x_{{0}}-X_{0}) }}( {\omega}^{2}-2\,\sqrt {1-{\omega}^{2}}-2 ) }{1+{{e}^{-2(\,x_{{0}}-\,X_{0})}}}}{dX_{0}}.\label{A4}
\end{eqnarray}


\begin{thebibliography}{lll}

\bibitem{bula77} L.N. Bulaevskii, V.V. Kuzii, and A.A. Sobyanin,
\emph{Superconducting system with weak coupling to the current in the ground state},
Pis'ma Zh. Eksp. Teor. fiz. {\bf25}, 314 (1977) [JETP Lett. {\bf 25}, 290 (1977)].

\bibitem{bula78} L.N. Bulaevskii, V.V. Kuzii, and A.A. Sobyanin,
\emph{On possibility of the spontaneous magnetic flux in a Josephson junction containing magnetic impurities},
Solid State Comm. {\bf 25}, 1053 (1978).


\bibitem{hilg08} H.\ Hilgenkamp, \emph{$\pi$-phase shift Josephson structures}, Supercond. Sci. Technol. {\bf 21}, 034011 (2008).

\bibitem{gold09} C.\ G\"urlich, E.\ Goldobin, R.\ Straub, D.\ Doenitz, Ariando, H.-J.H.\ Smilde, H.\ Hilgenkamp, R.\ Kleiner, D.\ Koelle, \emph{Imaging of order parameter induced $\pi$ phase shifts in cuprate superconductors by low-temperature scanning electron microscopy}, Phys. Rev. Lett. {\bf 103}, 067011 (2009).

\bibitem{pegr06} C. M. Pegrum, \emph{Can a fraction of a quantum be better than a whole one?}, Science {\bf 312}, 1483 (2006).

\bibitem{ortl06} T. Ortlepp, Ariando, O. Mielke, C. J. M. Verwijs, K. F. K. Foo, H. Rogalla, F. H. Uhlmann, and H. Hilgenkamp, \emph{Flip-flopping fractional flux quanta}, Science {\bf 312}, 1495-1497 (2006).

\bibitem{kato97} T. Kato and M. Imada, \emph{Vortices and quantum tunneling in current-biased $0-\pi-0$ Josephson junctions of d-wave superconductors},  J. Phys. Soc. Jpn. {\bf 66}, 1445 (1997).

\bibitem{gold05} E. Goldobin, H. Susanto, D. Koelle, R. Kleiner, and S. A. van Gils, \emph{Oscillatory eigenmodes and stability of one and two arbitrary fractional vortices in long Josephson $0-\kappa$ junctions},  Phys. Rev. B {\bf 71}, 104518 (2005).

\bibitem{ahma09} S. Ahmad, H. Susanto, J.A.D. Wattis, \emph{Existence and stability analysis of finite $0-\pi-0$ Josephson junctions}, Phys. Rev. B {\bf80}, 064515 (2009).

\bibitem{derk07} G. Derks, A. Doelman, S.A. van Gils, and H. Susanto, \emph{Stability analysis of $\pi$-kinks in a 0-$\pi$ Josephson junction},  SIAM J. Appl. Dyn. Syst. {\bf 6}, 99-141 (2007).

\bibitem{voge08} K. Vogel, T. Kato, W. P. Schleich, D. Koelle, R. Kleiner, and E. Goldobin, \emph{Theory of fractional vortex escape in a $0-\kappa$ long Josephson junction},  Phys. Rev. B {\bf 80}, 134515 (2009).

\bibitem{gold05_2} E. Goldobin, K. Vogel, O. Crasser, R. Walser, W. P. Schleich, D. Koelle, and R. Kleiner, \emph{Quantum tunneling of semifluxons in a $0-\pi-0$ long Josephson junction}, Phys. Rev. B.{\bf 72}, 054527 (2005).

\bibitem{buck07} K. Buckenmaier, T. Gaber, M. Siegel, D. Koelle, R. Kleiner, and E. Goldobin, \emph{Spectroscopy of the fractional vortex eigenfrequency in a long Josephson $0-\kappa$ junction}, Phys. Rev. Lett. {\bf 98}, 117006, (2007).

\bibitem{pfei09} J. Pfeiffer, T. Gaber, D. Koelle, R. Kleiner, E. Goldobin, M. Weides, H. Kohlstedt, J. Lisenfeld, A. K. Feofanov, and
A. V. Ustinov, \emph{Escape rate measurements and microwave spectroscopy of 0, $\pi$, and 0-$\pi$ ferromagnetic Josephson tunnel junctions},  arXiv:0903.1046. (2009).

\bibitem{gron04} N. Gr\o nbech-Jensen, M. G. Castellano, F. Chiarello, M. Cirillo, C. Cosmelli, L. V. Filippenko, R. Russo, and G. Torrioli, \emph{Microwave-induced thermal escape in Josephson junctions},  Phys. Rev. Lett. {\bf 93}, 107002 (2004).

\bibitem{gron04_2} N. Gr\o nbech-Jensen and M. Cirillo, \emph{AC-induced thermal vortex escape in magnetic-field-embedded long annular Josephson junctions },  Phys. Rev. B {\bf 70}, 214507 (2004).

\bibitem{guoz08} S. Guozhu, W. Yiwen, C. Junyu, C. Jian, J. Zhengming, K. Lin, X. Weiwei, Y. Yang, H. Siyuan, and W. Peiheng, \emph{Microwave-induced phase escape in a Josephson tunnel junction}, Phys. Rev. B {\bf 77}, 104531 (2008).

\bibitem{hilg03} H. Hilgenkamp, Ariando, H. J. H. Smilde, D.H.A. Blank, G. Rijnders, H. Rogalla, J.R. Kirtley, and C.C. Tsuei,
\emph{Ordering and manipulation of the magnetic moments in large-scale superconducting $\pi$-loop arrays},
Nature {\bf 422}, 50 (2003).

\bibitem{frol08} S.M. Frolov, M.J.A. Stoutimore, T.A. Crane, D.J. Van Harlingen, V.A. Oboznov, V.V. Ryazanov, A. Ruosi, C. Granata, M. Russo,
\emph{Imaging spontaneous currents in superconducting arrays of $\pi$-junctions}, Nature Physics {\bf 4}, 32-36 (2008).

\bibitem{oxto09} O. F. Oxtoby and I. V. Barashenkov, \emph{Wobbling kinks in $\phi^4$ theory},  Phys. Rev. E {\bf 80}, 026608 (2009).

\bibitem{oxto09_2} O. F. Oxtoby and I. V. Barashenkov, \emph{Resonantly driven wobbling kinks},  Phys. Rev. E {\bf 80}, 026609 (2009).

\bibitem{segu83} H. Segur, \emph{Wobbling kinks in $\phi^{4}$ and sine-Gordon theory }, J. Math. Phys. {\bf 24}, 1439 (1983).

\bibitem{kalb04} G. K\"albermann, \emph{The Sine-Gordon Wobble}, J. Phys. A {\bf 37}, 11603 (2004).

\bibitem{kevr00} P.G.\ Kevrekidis and M.I.\ Weinstein, \emph{Dynamics of lattice kinks}, Physica D \textbf{142}, 113-152 (2000).



\bibitem{fere08} L. A. Ferreira, B. Piette, and W. J. Zakrzewski, \emph{Wobbles and other kink-breather solutions of the sine-Gordon model},  Phys. Rev. E {\bf 77}, 036613 (2008).

\bibitem{dewe08} A. Dewes, T. Gaber, D. Koelle, R. Kleiner, and E. Goldobin, \emph{Semifluxon molecule under control},  Phys. Rev. Lett. {\bf 101}, 247001 (2008).

\bibitem{hans06} J.A. Boschker, \emph{Manipulation and on-chip readout of fractional flux quanta}, Master thesis, University of Twente, (2006).

\bibitem{gold04} E. Goldobin, A. Sterck, T. Gaber, D. Koelle, and R. Kleiner, \emph{Dynamics of Semifluxons in Nb long Josephson 0-$\pi$ junctions},  Phys. Rev. Lett. {\bf 92}, 057005 (2004).

\bibitem{goldo02} E. Goldobin, A. M. Klushin, M. Siegel, and N. Klein, \emph{Long Josephson junction embedded into a planar resonator at microwave frequencies: Numerical simulation of fluxon dynamics}, J. Appl. Phys. {\bf 92}, 3239-3250 (2002).

\bibitem{bark04} F. L. Barkov, M. V. Fistul, and A. V. Ustinov, \emph{Microwave-induced flow of vortices in long Josephson junctions}, Phys. Rev. B {\bf70}, 134515 (2004).


\bibitem{bamb09} D.\ Bambusi and S.\ Cuccagna, \emph{On dispersion of small energy solutions of the nonlinear Klein-Gordon equation with a potential}, arXiv:0908.4548.


\end{thebibliography}
\end{document}